\newif\ifanonymousversion
\anonymousversionfalse

\ifanonymousversion

\documentclass[compsoc,conference,a4paper,10pt,times]{IEEEtran}

\else

\documentclass[conference]{IEEEtran}

\fi

\usepackage{tikz}
\usepackage{amsmath}
\usepackage{xcolor}
\usepackage{hyperref}
\usepackage{soul}
\usepackage[braket, qm]{qcircuit}
\usepackage{physics}
\usepackage{enumitem}
\usepackage{multirow}
\usepackage{booktabs}
\usepackage{algorithmicx}
\usepackage{algpseudocode}
\usepackage{algorithm}
\usepackage{balance}
\usepackage{nicematrix}
\usepackage{subcaption}
\usepackage{amsfonts}
\usepackage{flushend}

\newcommand{\nsf}[1]{\href{https://www.nsf.gov/awardsearch/showAward?AWD_ID=#1}{#1}}

\begin{document}

\date{}

\title{Exploiting Reset Operations\\ in Cloud-based Quantum Computers\\ to Run Quantum Circuits for Free}

\ifanonymousversion

\else

\author{
\IEEEauthorblockN{Jakub Szefer}
\IEEEauthorblockA{\textit{Dept. of Electrical \& Computer Engineering} \\
\textit{Northwestern University}\\
Evanston, IL, USA \\
jakub.szefer@northwestern.edu}
}

\fi

\ifanonymousversion

\else

\IEEEoverridecommandlockouts
\makeatletter\def\@IEEEpubidpullup{6.5\baselineskip}\makeatother
\IEEEpubid{\parbox{\columnwidth}{
This work was supported in part by NSF grant \nsf{2332406}.
}
\hspace{\columnsep}\makebox[\columnwidth]{}}

\fi

\maketitle
\pagestyle{plain}

\begin{abstract}
This work presents the first thorough exploration of how  reset operations in cloud-based quantum computers could be exploited to run quantum circuits for free. This forms a new type of attack on the economics of cloud-based quantum computers. All major quantum computing companies today offer access to their hardware through some type of cloud-based service. Due to the noisy nature of quantum computers, a quantum circuit is run many times to collect the output statistics, and each run is called a shot. The fees users pay for access to the machines typically depend on the number of these shots of a quantum circuit that are executed.  Per-shot pricing is a clean and straightforward approach as users are charged a small fee for each shot of their circuit. This work demonstrates that per-shot pricing can be exploited to get circuits to run for free when users abuse recently implemented mid-circuit qubit measurement and reset operations. Through evaluation on real, cloud-based quantum computers this work shows how multiple circuits can be executed together within a shot, by separating each user circuit by set of reset operations and submitting all the circuits, and reset operations, as one larger circuit. As a result, the user is charged per-shot pricing, even though inside each shot are multiple circuits. Total per-shot cost to run certain circuits could be reduced by up to $900$\% using methods proposed in this work, leading to significant financial losses to quantum computing companies. To address this novel finding, this work proposes a clear approach for how users should be charged for their execution, while maintaining the flexibility and usability of the mid-circuit measurement and reset~operations.
\end{abstract}

\section{Introduction}
\label{sec:introduction}

Quantum computing has experienced rapid progress in recent years, transitioning from early prototypes with only one or two qubits to much larger systems. For example, quantum processors with up to $1,121$ qubits are now commercially accessible, and current technological roadmaps anticipate the development of modular, error-corrected quantum systems with approximately $200$ logical qubits capable of executing on the order of $100$ million quantum gates by the end of this decade~\cite{ibmqubit}. Present-day quantum devices fall under the Noisy Intermediate-Scale Quantum (NISQ) regime~\cite{preskill2018quantum}, characterized by limited qubit counts, lack of error correction~\cite{Devitt_2013}, and high levels of operational noise. Despite these limitations, NISQ devices have demonstrated potential utility in various domains, including combinatorial optimization, quantum chemistry, artificial intelligence, and financial modeling~\cite{lanyon2010towards,jones1998implementation,mermin2007quantum,biamonte2017quantum,farhi2014quantum,orus2019quantum,herman2023quantum}.

Contemporary quantum computing platforms are widely accessible via commercial cloud services such as IBM Quantum~\cite{ibm_quantum}, Amazon Braket~\cite{braket}, and Microsoft Azure~\cite{azure}. These platforms offer pay-as-you-go access models that are utilized not only by academic researchers but also by industry practitioners. As a result, many organizations, including startups and established enterprises, rely on cloud-based quantum infrastructure to execute proprietary quantum circuits without owning dedicated hardware. For instance, J.P. Morgan Chase employs the Quantinuum H-series quantum processor via cloud access to develop and test algorithms for solving linear systems on quantum hardware~\cite{yalovetzky2024solving}.

Quantum circuits such as developed by various companies or researchers serve as the foundational building blocks for quantum computation. These circuits encode quantum algorithms and enable the execution of complex computations, offering the potential to outperform classical computers in specific tasks. Typical quantum circuits are usually built using quantum gates, which are abstract units that define specific operations on qubits. In addition to gates, quantum circuits today have two other parts. Initially, all qubits need to start in a known state, typically the basis $\ket{0}$ state. Further, at the end of circuit execution, all the qubits are measured. Because of the noise in today's quantum computers, each circuit is executed many times, each execution is called a shot. Outputs from all the shots are combined together to collect the output probabilities for different states, which together form the output of the quantum circuit and algorithm it realizes. For example, for many quantum algorithms, the output state with the highest probability corresponds to the desired solution of the algorithm. Each execution of an algorithm or circuit can be abstracted as a task that is composed of many~shots.

\begin{figure*}[t]
    \centering
    \includegraphics[width=1\linewidth]{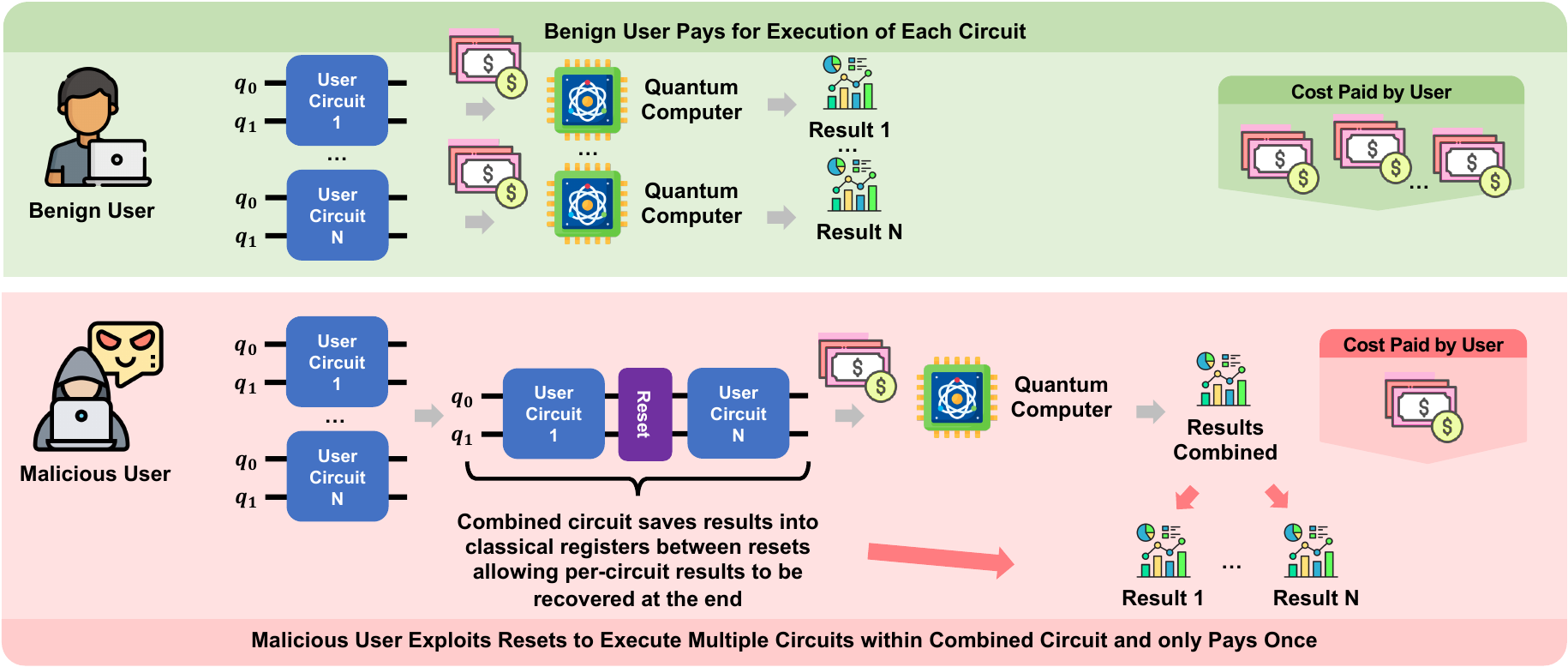}
    \caption{Overview of the proposed attack: the attacker subverts the billing strategies of cloud-based quantum computers (by abusing the reset quantum gate operations) to get free computation by running many quantum circuits for the cost of one quantum circuit.}
    \label{fig_free_shots_overview}
\end{figure*}

A recent feature of quantum computers are mid-circuit measurement and reset operations. The mid-circuit measurement and reset operation feature is important as it allows certain qubits to be reset while the larger circuit is executing. This allows for qubit reuse~\cite{decross2023qubit}, for example. In common applications of mid-circuit measurement and reset operations, only a few qubits are reset. Our proposed exploit takes the idea of resetting qubits to an extreme: reset all qubits within a circuit during execution. By resetting all qubits, the state of the quantum computer's qubits is set to $\ket{0}$, and new computation can be done on the qubits. To save the state of the qubits prior to the reset, mid-circuit measurement is used and data is saved in classical registers. An overview of this idea is shown in Figure~\ref{fig_free_shots_overview}.

Resetting all qubits are running a new computation can be made into an exploit in context of how cloud-based quantum computers charge for execution of the circuits. Because each circuit, i.e. task, is made up of many shots of the same set of gates, the various quantum computing platforms today employ a pricing model that combines both per-task and per-shot charges. In this model, users incur a fixed per-task fee for each submitted job, which covers compilation, queuing, and scheduling overheads. In addition, a per-shot fee is applied based on the number of circuit executions, i.e. shots, within the task. This structure ensures that providers can account for both the operational cost of job orchestration and the actual computational load imposed by quantum circuit execution. This dual-pricing strategy offers better cost recovery for providers and encourages users to optimize their circuit usage, but it also introduces additional complexity in budgeting and metering. From a security perspective, such a model can influence attack surfaces related to cost abuse, which is demonstrated in this~work. 

The particular new exploit presented in this work relates to the reset operation proposed, and now implemented, in number of quantum computers. As discussed earlier, the reset is a mechanisms to set qubits to a known basis state, typically $\ket{0}$. By actively resetting qubits within a shot, multiple circuits (or multiple copies of same circuit) can execute in a single shot. This effectively allows users to pay for fewer shots and fewer circuit execution fees. The demonstrations shown in this work demonstrate that total per-shot cost to run certain circuits could be reduced by up-to  $900$\%, leading to significant financial losses to quantum computing companies. To address this novel finding, this work proposes updated approach for how users should be charged for their execution, while maintaining the flexibility and usability of the mid-circuit measurement and active reset operations.

\section{Background}
\label{sec:background}

This section aims to offer background on how data is manipulated in quantum computers and a typical workflow for running quantum computing programs. It explains quantum gates, quantum circuits, and mid-circuit measurement and reset operations. The mid-circuit measurement and reset operations are the key features of the quantum circuits that are exploited in this work.

\subsection{Qubits and Quantum States}

In quantum computing, the basic unit of information is the quantum bit (qubit), which serves a role analogous to that of the classical bit in conventional computation. A qubit resides in a two-dimensional Hilbert space and has two computational basis states, denoted using Dirac notation as $\ket{0}$ and $\ket{1}$. However, unlike classical bits which are constrained to values of either 0 or 1, a qubit can exist in a superposition of these states: $\ket{\psi} = \alpha \ket{0} + \beta \ket{1}$, where $\alpha, \beta$ are complex and $|\alpha|^2 + |\beta|^2 = 1$.

Qubits are often represented using vector representation. For instance, the single-qubit basis states can be denoted as two-dimensional column vectors: $\ket{0} = \begin{bmatrix} 1, \ 0 \end{bmatrix}^T$ and $\ket{1} = \begin{bmatrix} 0, \ 1 \end{bmatrix}^T$, where $T$ is the transpose. Thus, the state $\ket{\psi}$ can be expressed as $\ket{\psi} = \alpha \ket{0} + \beta \ket{1} = \begin{bmatrix} \alpha, \ \beta \end{bmatrix}^T$. Extending this notion to $n$ qubits, the space of $n$-qubit states encompasses $2^n$ basis states, ranging from $\ket{0\dots 0}$ to $\ket{1\dots 1}$. Consequently, an $n$-qubit state $\ket{\phi}$ can be represented as: $\ket{\phi} = \sum_{i = 0}^{2^n - 1} a_i \ket{i}$, where $\sum_{i = 0}^{2^n - 1}|a_i|^2 = 1$.

\subsection{Quantum Gates}

In quantum computing, fundamental operations are implemented as quantum gates, which apply unitary transformations to qubits. Quantum circuits execute sequences of such gates to coherently evolve qubit states toward a desired computational outcome.

A quantum gate $U$ must be unitary, i.e., $U U^\dagger = U^\dagger U = I$, where $U^\dagger$ signifies the conjugate transpose of $U$, and $I$ is the identity matrix. Operating on a qubit $\ket{\psi}$, a quantum gate $U$ transforms it as $\ket{\psi} \rightarrow U \ket{\psi}$. Employing the matrix representation, $n$-qubit quantum gates are expressed as $2^n \times 2^n$ matrices. Quantum gates can be classified as single-qubit gates or multi-qubit gates. U3 gate is one general single-qubit gate that is parameterized by 3 Euler angles. Some other single-qubit gates include Pauli-$X$ gate: a single-qubit gate similar to the classical NOT gate, flips $\ket{0}$ to $\ket{1}$ and vice versa; {\tt RZ} gate: a phase shift between $\ket{0}$ and $\ket{1}$. Unlike classical computing, multi-qubit gates can create quantum entanglement. One notable example is the CNOT gate, also known as the CX gate, a two-qubit gate that applies a Pauli-$X$ gate to the target qubit if the control qubit is in state $\ket{1}$, otherwise, it remains unchanged. Another example is the control-U3 gate, which also uses one bit to control the application of the U3 gate.

Matrix representations of some gates are presented below, adhering to the qubit order specified by Qiskit~\cite{Qiskit}:

\begin{equation*}
{\tt U3(\theta, \phi, \lambda)}=\begin{bmatrix}
\cos(\frac{\theta}{2}) & -e^{i\lambda} \sin(\frac{\theta}{2}) \\
e^{i\phi} \sin(\frac{\theta}{2}) & e^{i(\phi + \lambda)} \cos(\frac{\theta}{2})
\end{bmatrix}
\end{equation*}
\begin{equation*}
{\tt RZ(\theta)}=\begin{bmatrix}
e^{-i\frac{\theta}{2}} & 0 \\
0 & e^{i\frac{\theta}{2}}
\end{bmatrix},
{\tt CX} = \begin{bmatrix}
1 & 0 & 0 & 0 \\
0 & 0 & 0 & 1 \\
0 & 0 & 1 & 0 \\
0 & 1 & 0 & 0
\end{bmatrix}
\end{equation*}

It has been shown that any unitary quantum operation can be efficiently approximated, up to an arbitrary precision, using a finite set of quantum gates~\cite{deutsch1995universality}. As a result, modern quantum processors typically support a restricted set of basis, or native, gates. By composing these native gates, more complex operations can be synthesized, enabling the emulation of arbitrary quantum gates within a given hardware architecture.

\subsection{Quantum Circuits}

A quantum circuit consists of a sequence of quantum gates applied to qubits, followed by measurement. After constructing a high-level circuit using a quantum software development framework such as Qiskit, the circuit must be transpiled into hardware-specific instructions. Transpilation maps the circuit to a set of basis gates compatible with the target hardware's native gate set and topology. For superconducting qubits, gate operations are implemented via microwave pulses, which are generated during the scheduling stage, wherein gate-level instructions are translated into pulse-level commands. Upon completion of all gate operations, the qubits are measured. The resulting measurement probabilities encode the solution of the circuit. A quantum circuit operating on $n$ qubits yields $2^n$ potential measurement outcomes. However, due to inherent noise in current Noisy Intermediate-Scale Quantum (NISQ) devices, each circuit must be executed thousands of times—each execution referred to as a shot—to obtain statistically meaningful probability estimates. The final computational result is inferred from these probabilities.

\subsection{Mid-Circuit Measurements and Reset Operations in Quantum Circuits}

Quantum computation often requires qubits to be initialized to a known basis state, typically the $\ket{0}$ state, prior to circuit execution. While passive reset (i.e., waiting for qubits to naturally relax to the ground state via thermalization and decoherence) is straightforward, it incurs latency on the order of microseconds to milliseconds, depending on the qubit technology. To address this inefficiency, many quantum platforms implement active reset operations, a technique that deterministically prepares qubits in a desired state using measurement and conditional feedback operations~\cite{egger2018pulsed}.

In active reset, a qubit is first measured, and if found in the excited state $\ket{1}$, a corrective gate (e.g., an $X$ gate) is applied to return it to $\ket{0}$. This approach significantly reduces the initialization time, often to under a microsecond, enabling faster circuit execution and higher throughput in quantum workloads. Active reset is especially critical in algorithms requiring mid-circuit measurement and reuse of qubits (e.g., in quantum error correction, teleportation protocols, or certain variational algorithms).

Active reset is closely related to mid-circuit measurement. Mid-circuit measurement in quantum computing refers to the ability to measure a qubit during the execution of a quantum circuit, rather than only at the end. This enables dynamic behavior and conditional logic based on quantum outcomes in real time, such as application of the $X$ gate to flip the qubit state to reset the qubit. As part of mid-circuit measurement, the measurement results are saved into classical registers. Multiple mid-circuit measurements (and resets) can be applied to the same qubit, each time saving the results into a different classical register. Thus, data from measurement can be saved into classical register $c_i$, then the qubit is reset, some more operations are performed, it is measured again saving data into classical register $c_{i+1}$, etc. All the classical registers are returned to the user at end of circuit execution.

\section{Threat Model}
\label{sec:threat_model}

This work follows a very simple threat model: we assume that a malicious user has only regular, user-level access to quantum computers which implement mid-circuit measurement and active reset. Today, mid-circuit measurement and active reset are either supported, or have been announced for quantum computers such as from IBM, Rigetti, or IQM. We assume the malicious user is able to compile their quantum circuits locally and concatenate multiple circuits (each separated by a set of reset gates) into a larger circuit submitted to the quantum computer. Common quantum computing programming frameworks such as Qiskit allow users to do this with no special privileges nor modifications. We assume that billing is done on per-task plus per-shot basis, or on a similar time-based basis where fixed per-task cost dominates and user is effectively charged less, or even much less, if they have one large task (made of the concatenated circuits). We assume there is no mechanism used by cloud providers for detection of this abuse of reset operation, as is the case~today.

\section{Cloud-Based Quantum Computing Pricing}
\label{sec_pricing}

Basis of our exploit is combination of reset operation and how it affects the billing of user circuits. In this section we discuss the different billing practices of cloud providers. Different cloud-based quantum computing providers offer different pricing models for access to quantum computers. We identify three models: per-task + per-shot, time-based (implicit per-task + per-shot) and per-gate execution.

\subsection{Per-Task + Per-Shot Pricing}

Major cloud providers such as AWS and their AWS Braket service use explicit er-task + per-shot pricing. Table~\ref{tab_asw_braket_pricing} shows the price for each of the different quantum computers, also called quantum processing units (QPUs) available from AWS Braket%
\footnote{Source: \url{https://aws.amazon.com/braket/pricing/}, data retrieved on June 1,~2025.}. Fixed per-task cost is assumed to cover the fixed costs such as compilation, and per-shot covers the time spent on the quantum hardware. User is charged per-task, plus per-shot costs for each shot of their job.

If a malicious user is able to combine $C$ different circuits each executing for $S$ shots into one task, they can avoid paying per-task fees for each circuit; and while they pay per-shot cost, each shot includes $N$ circuits, so the effective cost they avoid paying is $(N-1) \times S \times pershotcost$.

\begin{table}[t]
\centering
\caption{Quantum hardware pricing for different providers and QPUs from AWS Braket.}
\begin{tabular}{lllll}
\toprule
\textbf{Provider} & \textbf{QPU} & \textbf{Per-task Price} & \textbf{Per-shot Price} \\
\midrule
IonQ     & Forte  & \$0.30000 & \$0.08000 \\
IonQ     & Aria   & \$0.30000 & \$0.03000 \\
IQM      & Garnet & \$0.30000 & \$0.00145 \\
QuEra    & Aquila & \$0.30000 & \$0.01000 \\
Rigetti  & Ankaa-3  & \$0.30000 & \$0.00090 \\
\bottomrule
\end{tabular}
\label{tab_asw_braket_pricing}
\end{table}

\begin{table}[t]
\centering
\caption{Assessment of vulnerability and testing of different quantum billing models in this work.}
\label{tab_billing_model_vuln}
\begin{tabular}{lcc}
\toprule
\textbf{Billing} & \textbf{Possibly} & \textbf{Demonstrated Attack} \\
\textbf{Model} & \textbf{Vulnerable} & \textbf{in This Work} \\
\midrule
Per-Task + Per-Shot & Yes & -- \\
Time-based & Yes & Yes \\
Per-Gate Execution & No & -- \\
\bottomrule
\end{tabular}
\end{table}

\subsection{Time-based Pricing}

\begin{figure}[t]
     \centering
     \includegraphics[width=0.98\columnwidth]{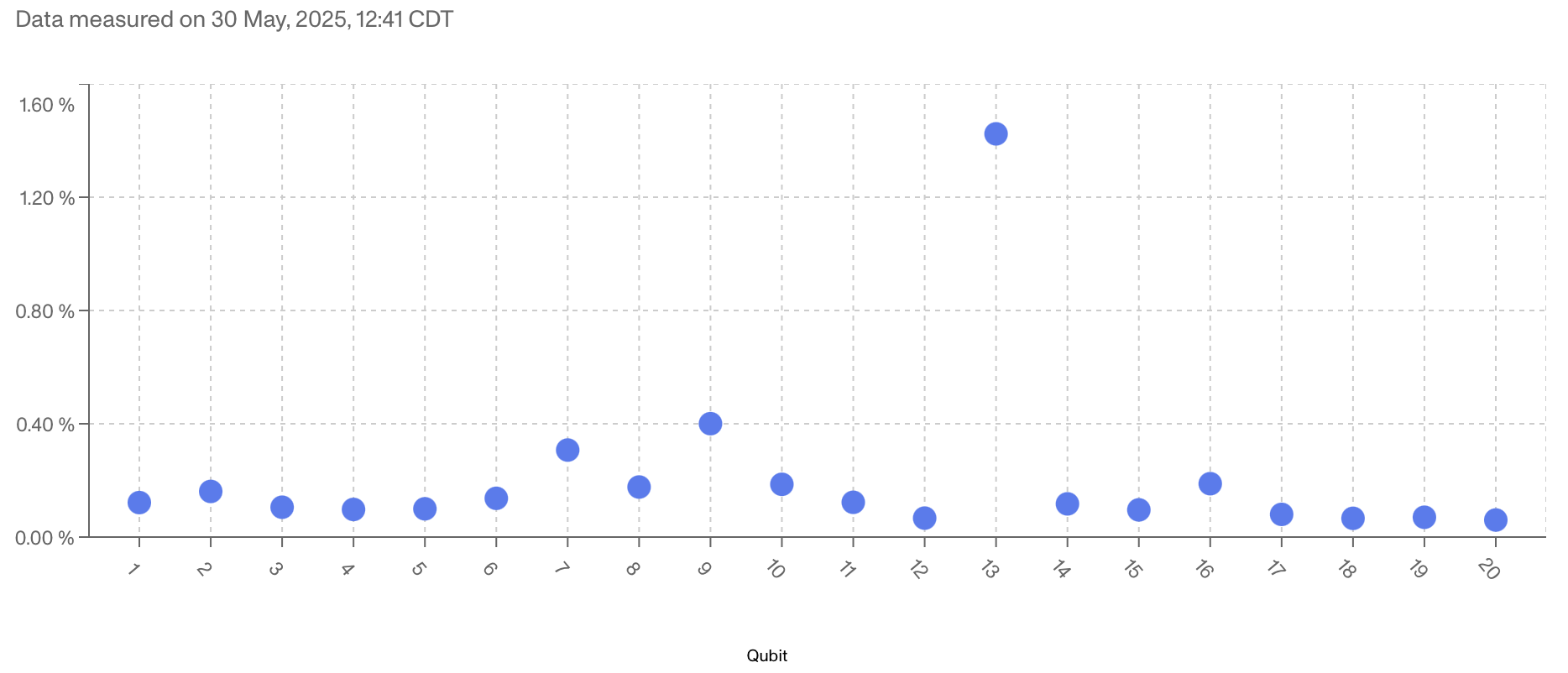}
     \caption{\small Conditional $X$ gate error rates, data from the backend used in the evaluation.}
    \label{fig_prx_gate_error}
\end{figure}

\begin{figure}[t]
     \centering
     \includegraphics[width=0.98\columnwidth]{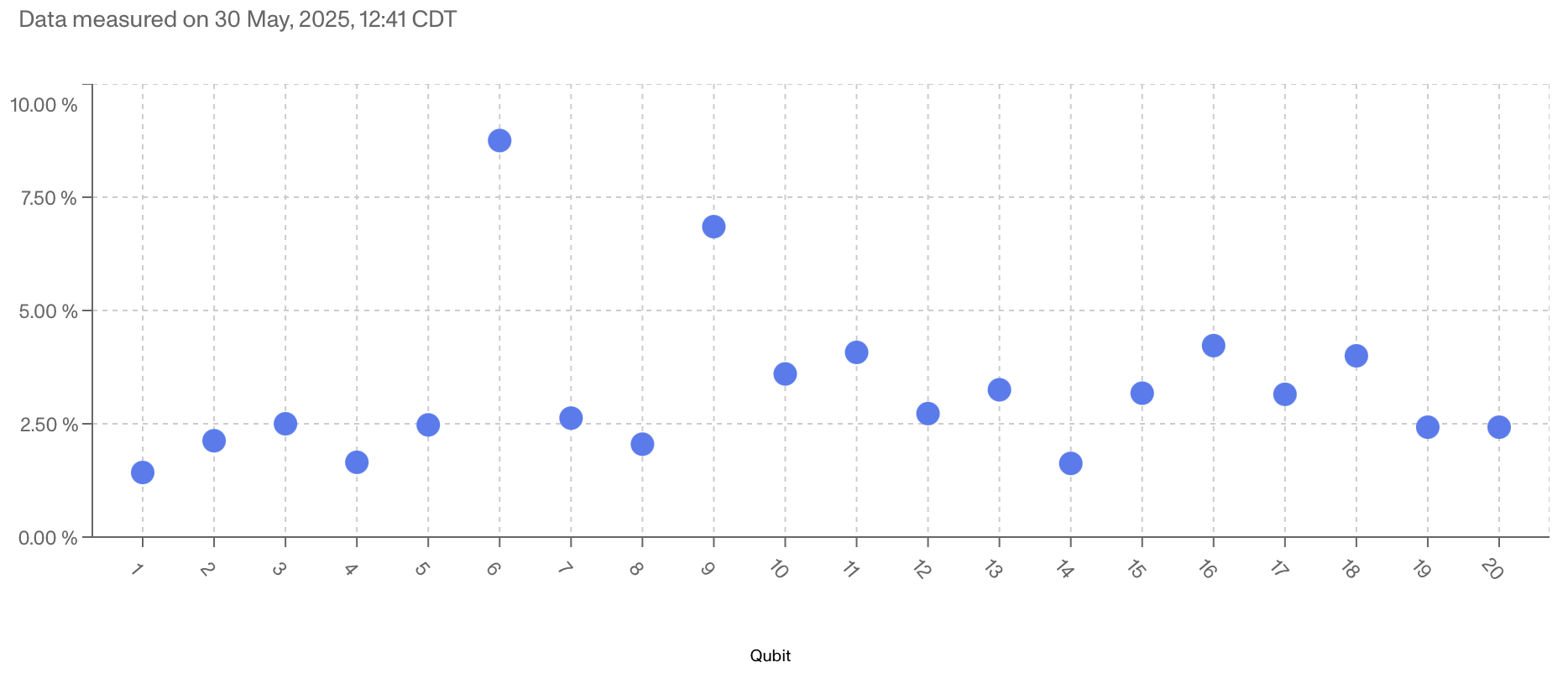}
     \caption{\small Readout error rates, data from the backend used in the evaluation.}
    \label{fig_readout_error}
\end{figure}

Other cloud providers charge time-based fees for using the quantum computers. Although it is called time-based pricing by the vendors, it is really an implicit per-task + per-shot pricing. Especially, the time-based fees include time spent compiling, setting up, and executing the circuit. Some quantum computers offered through Microsoft Azure follow this model. According to data retrieved from Azure%
\footnote{Source: \url{https://learn.microsoft.com/en-us/azure/quantum/pricing}, data retrieved on June 1,~2025.}, PASQAL Fresnel quantum computer use costs USD 0.08333 per second + Azure infrastructure costs, while Rigetti Ankaa-3 use costs USD 1.30 per second increment of job execution time. Quantinuum and IonQ available from Microsoft Azure follow per-gate model discussed later. Meanwhile, IBM charges USD 1.60 per second for their devices according to their data%
\footnote{Source: \url{https://www.ibm.com/quantum/pricing}, data retrieved on June 1,~2025.}. Yet different quantum computer provider, IQM charges USD 0.30 per second according to their data%
\footnote{Source: \url{https://meetiqm.com/products/iqm-resonance/} data retrieved on June 1,~2025.}.

\subsection{Per-Gate Execution Pricing}

The last billing model is by IonQ and Quantinuum through the Microsoft Azure service%
\footnote{Source: \url{https://learn.microsoft.com/en-us/azure/quantum/pricing}, data retrieved on June 1,~2025.}. For these machines, the users are charged per gate (single qubit or two-qubit gate). Thus longer circuits, including ones with resets in the middle of them, will be charged more.

\subsection{Which Billing Models Can be Exploited}

Table~\ref{tab_billing_model_vuln} shows the different billing models. Our access to one of the quantum providers and machines with the time-based billing model has allowed us to demonstrate exploitation of reset operations in cloud-based quantum computers to run quantum circuits for free. Demonstration of attack for per-task plus per-shot model is left as future work as it involves different types of quantum computing hardware. Meanwhile, we believe per-gate execution model is not vulnerable if the per-gate billing is correctly implemented.

\section{Reset Operations in Quantum Computers}

\begin{figure*}[t]
     \centering
     \includegraphics[width=1\textwidth]{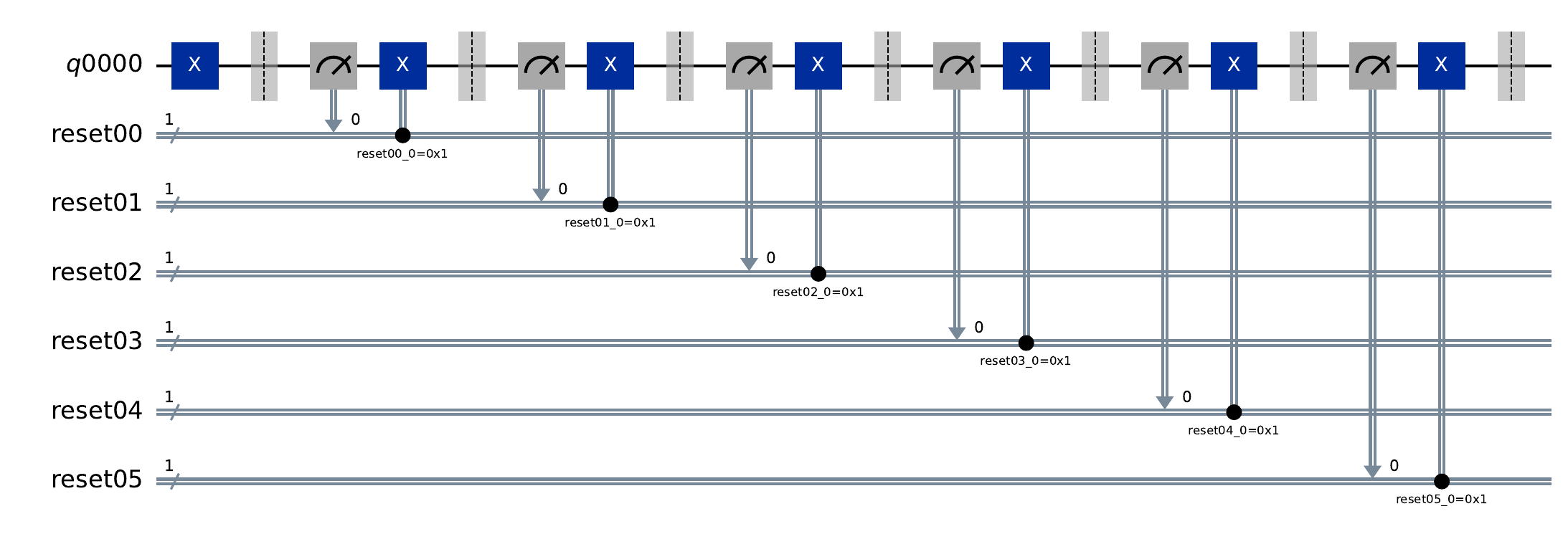}
     \caption{\small Illustration circuit used to test the fidelity of different number of resets. Due to space limitation, only example with 6 resets is shown in this figure.}
    \label{fig_reset_test}
\end{figure*}

We now introduce reset operations in quantum computers. The data and description is based on the target quantum computer evaluated in this work. Note Section~\ref{sec_disclosure} discusses our responsible disclosure practice undertaken prior to submission of this work.

\subsection{Reset Gate}

In the various quantum computers, the reset gate follows effectively the same design. A qubit is measured. Measurement of the qubit collapses the quantum state to either $\ket{0}$ or $\ket{1}$ state, i.e. classical $0$ or classical $1$. All quantum state is lost after a measurement. However, the qubit could be in the $\ket{1}$ state. In that case, a conditional $X$ gate is applied to the qubit. A quantum $X$ gate is similar to classical $NOT$ operation: it exchanges amplitudes of the basis states of the qubit. If the qubit is in $\ket{1}$ state, the application of $X$ gate will put the qubit into $\ket{0}$ gate. Thus a reset is a measurement followed by conditional $X$ gate ensuring the qubit is in $\ket{0}$ state after the reset is done. However, in practice, the reset is not perfect.

\subsection{Gate Error Rates}

Quantum gates, including measurement and $X$ gates, suffer from errors. Measurement gates suffer from readout error, with example data from our target quantum computer shown in Figure~\ref{fig_readout_error}. While the errors for the conditional $X$ gate are shown in Figure~\ref{fig_prx_gate_error}. Based on this, a qubit may need to be reset multiple times to get as close to $\ket{0}$ as possible.

\subsection{Evaluation of Ideal Reset Strategy}
\label{sec:attack}

In this section, we present our evaluation used to establish the number of reset operations needed to effectively set qubit into $\ket{0}$ state. This number of resets will later be used
between circuits within a same shot to effectively reset the qubits when our exploit is demonstrated. 

As mentioned above, each reset operation is not perfect. According to the public documentation for the backend used in the evaluation, the fidelity of the conditional $X$ gate is about $99.80$\%. In addition the readout fidelity is about $96.74$\%. An active gate requires both a measurement (readout) to read the qubit and then conditionally apply the $X$ gate. Based on the above public data, we would expect the effective reset fidelity to be about $96.54$\%. This equals error rate of $0.0346$ or $34$ shots out of a $1000$ should not be reset correctly. We want to establish the reset fidelity in practice, and also check how use of multiple resets can help improve the fidelity.

\subsection{Reset Fidelity}
\label{sec_eval_num_reset}

Figure~\ref{fig_reset_test} shows the test circuit used to evaluate the fidelity of the resets based on resetting a qubit when it is in $\ket{1}$ state. To test this, qubits tested are initially set to $\ket{1}$ state by applying an $X$ gate at the beginning of the circuit. Then each active operations consists of a barrier, measurement of the qubit, and the conditional $X$. The barriers have no practical effect and are only used to more clearly separate each reset for visual inspection.

After each reset, we analyze and plot the probability of the qubit being in the $\ket{0}$ basis state. With ideal reset, the qubit should always be in $\ket{0}$ after each reset. However, due to the imperfections in the reset, this is not always the case.

\begin{figure}[t]
    \centering
    \includegraphics[width=0.99\linewidth]{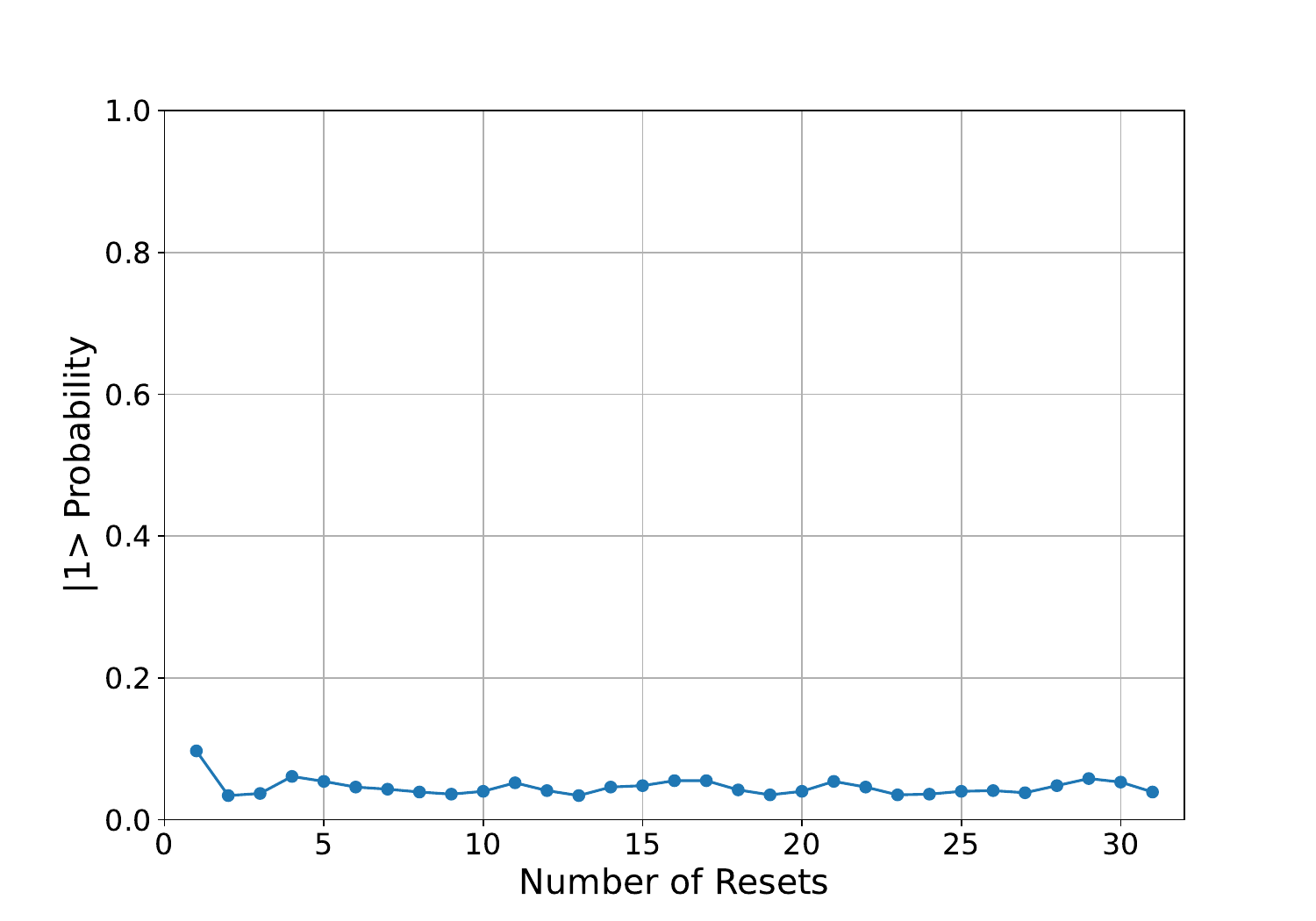}
    \caption{Probability of qubit being in $\ket{1}$ state for different number of resets based on the testing circuit from Figure~\ref{fig_reset_test}. Lower is better.}
    \label{fig_reset_test_result}
\end{figure}

Figure~\ref{fig_reset_test_result} shows the results of the tests. We tested from $1$ to $31$ resets. We execute the test circuit from Figure~\ref{fig_reset_test} for $1000$ shots. Note the figure only shows $6$ resets, while we test up to $31$. Based on the figure it can be observed that increasing number of resets beyond 1 helps to improve the fidelity (lower probability of qubit being in $\ket{1}$). However, due to noise and fluctuations in the error rates, going beyond 2 or 3 resets does not really have practical benefit considering the tested backend.

\subsection{Reset Strategy Confirmation using Bell State Circuit}

\begin{figure}[t]
    \centering
    \includegraphics[width=0.7\linewidth]{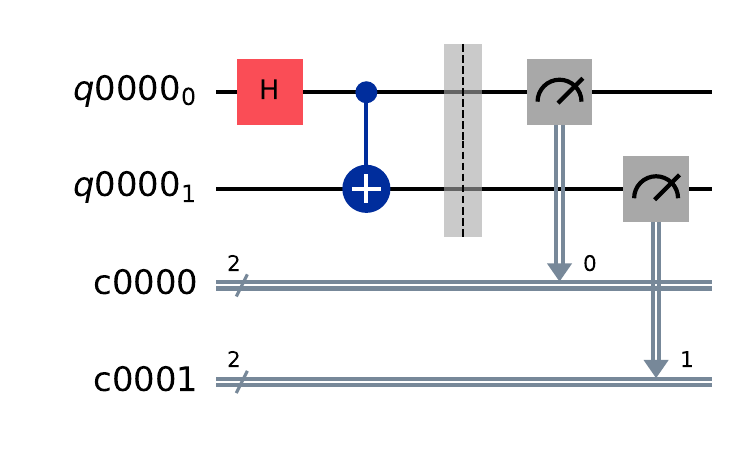}
    \caption{Bell state circuit example. The Bell state circuit uses a Hadamard $H$ gate followed by a $CNOT$ gate to generate entanglement between two qubits. The Hadamard gate creates a superposition on the first qubit, and the $CNOT$ gate entangles it with the second qubit by flipping its state conditionally. At the end of the circuit, the two qubits are measured. A barrier is shown to visually separate the circuit from measurement, the barrier performs no operation.}
    \label{fig_bell_state_circ}
\end{figure}

To validate the findings from Section~\ref{sec_eval_num_reset}, we tested a simple Bell state circuit. Bell State circuit is used to generate Bell states. Bell states are maximally entangled two-qubit quantum states that exhibit correlations stronger than those possible in classical systems. A Bell state represents a superposition in which the measurement outcomes of the two qubits are perfectly correlated, regardless of spatial separation. Bell states serve as foundational resources in quantum cryptographic protocols, such as quantum key distribution (QKD), and play a critical role in demonstrating quantum non-locality and designing secure quantum communication systems. Example of Bell state circuit is shown in Figure~\ref{fig_bell_state_circ}. In an ideal case, the output of a Bell state circuit should be $\ket{00}$ with $50$\% probability and $\ket{11}$ with $50$\% probability. Because the qubits are entangled, they should always be either both $0$ or both $1$, thus at the output of the circuit, states $\ket{01}$ and $\ket{10}$ should never occur.

Incorrect output, i.e. some occurrence of $\ket{01}$ or $\ket{10}$ states can happen for two reasons. First, the quantum gates and quantum computers in general are noisy and not perfect. This is something we cannot control and even with qubits starting in the ideal $\ket{0}$ states, we will never get $\ket{00}$ with $50$\% probability and $\ket{11}$ with $50$\% probability. Second, if qubits do not start in the ideal $\ket{0}$ states, then the output will further diverge from $\ket{00}$ with $50$\% probability and $\ket{11}$ with $50$\% probability. When we use active reset to reset qubits, this is something we can try to control by executing multiple resets to try to get qubits into $\ket{0}$ states more reliably.

\begin{figure}[t]
    \centering
    \includegraphics[width=1\linewidth]{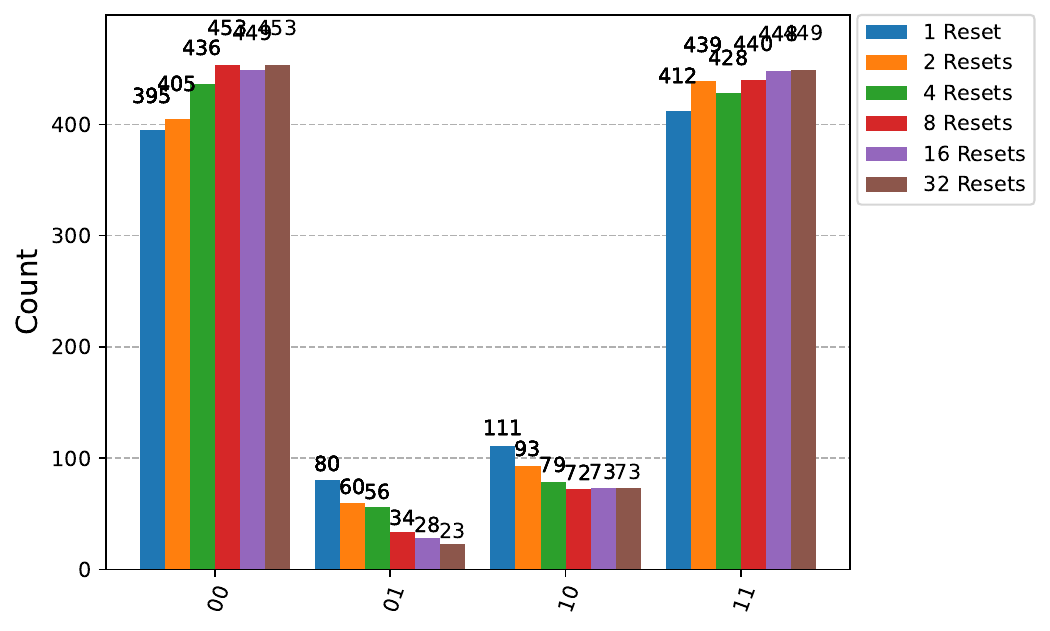}
    \caption{Output counts for Bell state circuit for different number of resets used between each copy of the circuit.}
    \label{fig_bell_state_num_resets_histogram}
\end{figure}

Based on results from Section~\ref{sec_eval_num_reset} about $4$ resets work ideally for getting qubits as close as possible to $\ket{0}$ state. To validate this, we run $8$ copies of bell state circuit within a shot, each copy being separated by either, 1, 2, 4, 8, or 16 resets. The whole circuit is executed for $1000$ shots. From the execution, we collect output statistics for the $8$ copies of Bell state circuit. We plot the output probabilities in Figure~\ref{fig_bell_state_num_resets_histogram}.

\begin{table}[t]
\centering
\caption{Total variational distance (TVD) between output probabilities for Bell-state circuits with different numbers of resets. TVD is computed relative to the baseline where only one circuit is executed per shot (1 copy). Lower is better.}
\label{tab:bell_state_num_resets_tvd}
\begin{tabular}{lc}
\toprule
\textbf{Number of Resets} & \textbf{TVD} \\
\midrule
2  & 0.0718 \\
4  & 0.0416 \\
8  & 0.0315 \\
16 & 0.0340 \\
32 & 0.0299 \\
\bottomrule
\end{tabular}
\end{table}

\begin{table*}[t]
\centering
\caption{Baseline cost for execution of Bell state circuit.}
\label{tab_bell_state_baseline_cost}
\begin{tabular}{ccccccc}
\toprule
\textbf{Circuit} & \textbf{Num. Qubits} & \textbf{Depth} & \textbf{Shots} & \textbf{Time (s)} & \textbf{Cost (credits)} & \textbf{Cost per Shot}\\
\midrule
Bell & 2 & 2 & 1000 & 2 & 1.5 & 0.001500\\
\bottomrule
\end{tabular}
\end{table*}

\begin{table*}[t]
\centering
\caption{Execution metrics for a Bell state circuit across varying repetition levels, with savings relative to baseline cost. The used is the Bell state circuit from Table~\ref{tab_bell_state_baseline_cost}, which uses $2$ qubits and has depth of $2$.}
\label{tab_bell_cost_scaling}
\begin{tabular}{ccccccccc}
\toprule
\textbf{Circuit} & \textbf{Reps} & \textbf{Resets} & \textbf{Shots} & \textbf{Eff. Shots} & \textbf{Time (s)} & \textbf{Cost (credits)} & \textbf{Cost per Shot} & \textbf{Savings} \\
\midrule
Bell & 64 & 4 & 1000 & 64000 & 14 & 10.5 & 0.000164 & 814.29\% \\
Bell & 32 & 4 & 1000 & 32000 & 6 & 4.5 & 0.000141 & 966.67\% \\
Bell & 16 & 4 & 1000 & 16000 & 4 & 3 & 0.000188 & 700.00\% \\
Bell & 8 & 4 & 1000 & 8000 & 3 & 2.25 & 0.000281 & 433.33\% \\
Bell & 4 & 4 & 1000 & 4000 & 3 & 2.25 & 0.000563 & 166.67\% \\
Bell & 2 & 4 & 1000 & 2000 & 2 & 1.5 & 0.000750 & 100.00\% \\
Bell & 1 & 4 & 1000 & 1000 & 2 & 1.5 & 0.001500 & 0.00\% \\
\bottomrule
\end{tabular}
\end{table*}

\begin{table*}[t]
\centering
\caption{Cost scaling with varying number of resets used between execution of $8$ copies of the Bell state circuit.}
\label{tab_reset_scaling}
\begin{tabular}{c|c|c|c|c|c|c|c|c}
\toprule
\textbf{Circuit} & \textbf{Reps} & \textbf{Resets} & \textbf{Shots} & \textbf{Eff. Shots} & \textbf{Time (s)} & \textbf{Cost (credits)} & \textbf{Cost per Shot} & \textbf{Savings} \\
\midrule
Bell & 8 & 1 & 1000 & 8000 & 3 & 2.25 & 0.000281 & 433.33\% \\
Bell & 8 & 2 & 1000 & 8000 & 3 & 2.25 & 0.000281 & 433.33\% \\
Bell & 8 & 4 & 1000 & 8000 & 3 & 2.25 & 0.000281 & 433.33\% \\
Bell & 8 & 8 & 1000 & 8000 & 3 & 2.25 & 0.000281 & 433.33\% \\
Bell & 8 & 8 & 1000 & 8000 & 4 & 3.00 & 0.000375 & 300.00\% \\
Bell & 8 & 8 & 1000 & 8000 & 4 & 3.00 & 0.000375 & 300.00\% \\
Bell & 8 & 16 & 1000 & 8000 & 6 & 4.50 & 0.000563 & 166.67\% \\
Bell & 8 & 32 & 1000 & 8000 & 14 & 10.50 & 0.001313 & 14.29\% \\
\bottomrule
\end{tabular}
\end{table*}

We also compute the Total Variational Distance (TVD) between the output probabilities for the Bell state circuits for different number of resets. The TVD is computed relative to Bell state circuit output when only $1$ copy is run, i.e. only one circuit is executed in a shot. We observe that with $4$ resets we achieve a low value. This correlates with experiments in Section~\ref{sec_eval_num_reset}.

\section{Executing Circuits for Free: Bell State Circuit Example}

In this section we continue with the Bell state circuit as our example to evaluate how putting multiple copies of a circuit into a job can allow user to effectively obtain computation for free. The baseline for discussion in this section is execution of one Bell state circuit for $1000$ shots. Table~\ref{tab_bell_state_baseline_cost} shows the parameters of the circuit, as well as the time charged for execution of the circuit and the cost (credits) for the target backend used in our work.

\subsection{Cost Compared to Multiple Separate Jobs}

When users submits a job, they are charged the per-task and per-shot costs. In case of the Bell state circuit, one job of $1000$ shots costs $1.5$ credits or $0.001500$ per shot on our target quantum computer. If user has to re-run the experiment multiple times, say $32$ times, they would have to pay $32*1.5=48$ credits for $32$ jobs, and have same $0.001500$ per shot cost.

Now, we show that by abusing the mid-circuit measurement and reset, user can get the same computation done for $966$\% less and pay only $4.5$ credits vs. $48$ if each copy of the circuit is executed separately. This means the user is able to get $43.5$ credits worth of computation for free for a sample Bell circuit.

In Table~\ref{tab_bell_cost_scaling} we show details of the cost breakdown for different number of repetitions of the circuit. Here $1$ repetition means equivalent of executing the circuit one time. I.e. if user needs to run their circuit $32$ times as discussed above, that is equal to $32$ repetitions.

In generating data for the Table~\ref{tab_bell_cost_scaling} we use $4$ resets between each copy of the circuit to ensure that qubits are property reset. Later we evaluate the cost vs. number of resets used. Of course, less resets reduce cost but decrease fidelity of the circuits, while more resets would increase the fidelity of the circuits but make the cost savings for the user less.

In the Table, Effective Shots column lists the effective shots the user was able to execute. For example, if user submits the job with $1000$ shots, but within each shot the circuit is executed $32$ times, i.e. Reps equals $32$, then in effect the user can obtain data for $32,000$ shots as they get measurement data for each of the $32$ copies of the circuit for each shot in their job of $1000$ shots.

\subsection{Cost as Function of Number of Resets Used}

Varying the number of resets between each copy of the circuit can be used to control fidelity vs. total cost. Fewer resets give worst fidelity, but the cost benefit is more. More resets give better fidelity, but the cost benefit is less.

\begin{figure*}[t]
    \centering

    \subfloat[Bernstein-Vazirani]{\includegraphics[width=0.2\textwidth]{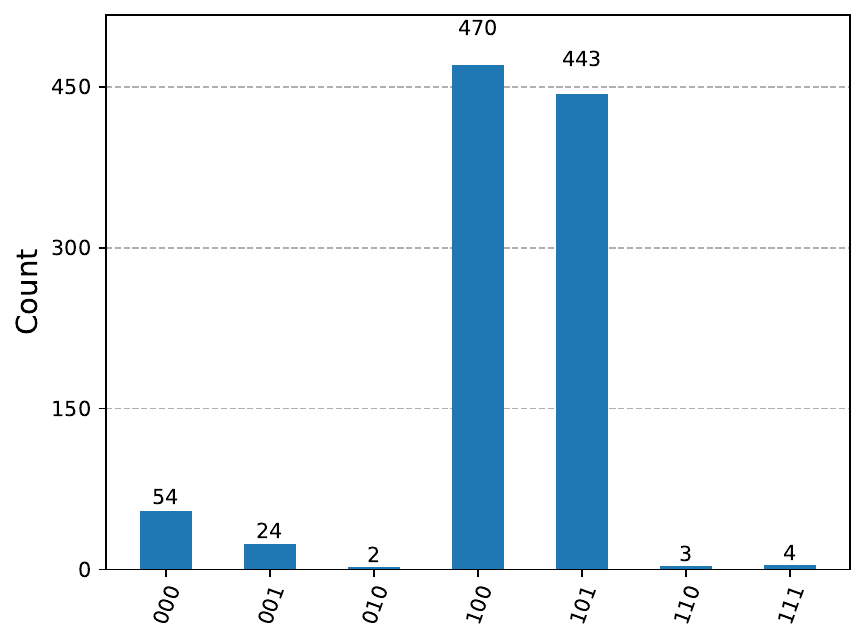}\label{fig:bernstein_vazirani}}
    \hfill
    \subfloat[Deutsch-Jozsa]{\includegraphics[width=0.2\textwidth]{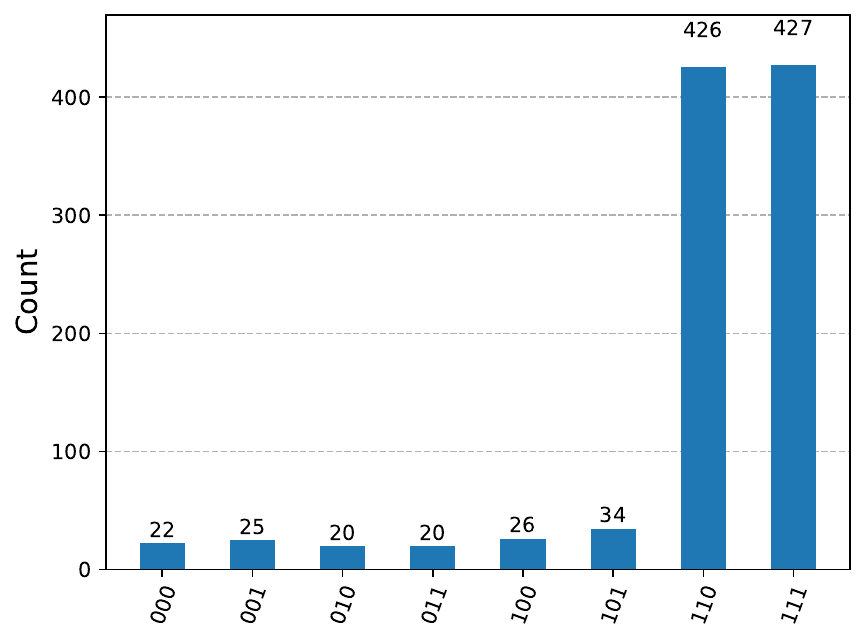}\label{fig:deutsch_jozsa}}
    \hfill
    \subfloat[Phase Estimation]{\includegraphics[width=0.2\textwidth]{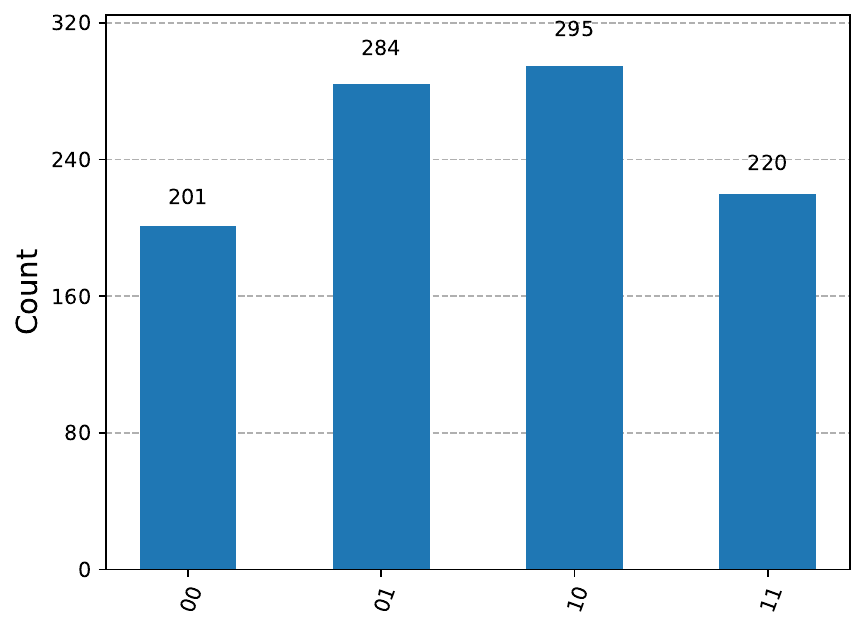}\label{fig:phase_estimation}}
    \hfill
    \subfloat[Grover's Oracle]{\includegraphics[width=0.2\textwidth]{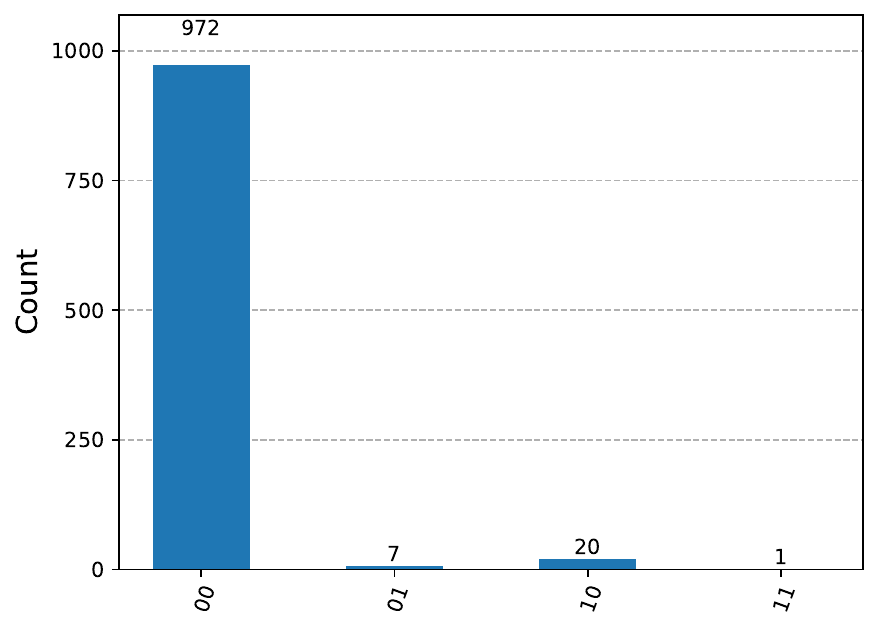}\label{fig:grover_oracle}}

    \vspace{0.3cm}
        
    \subfloat[Variational Ansatz]{\includegraphics[width=0.2\textwidth]{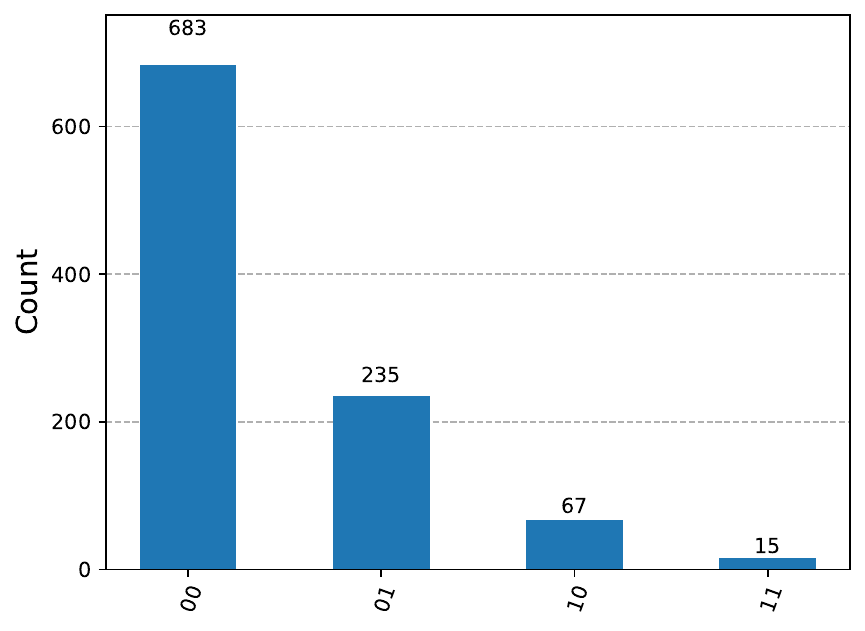}\label{fig:variational_ansatz}}
    \hfill
    \subfloat[Teleportation]{\includegraphics[width=0.2\textwidth]{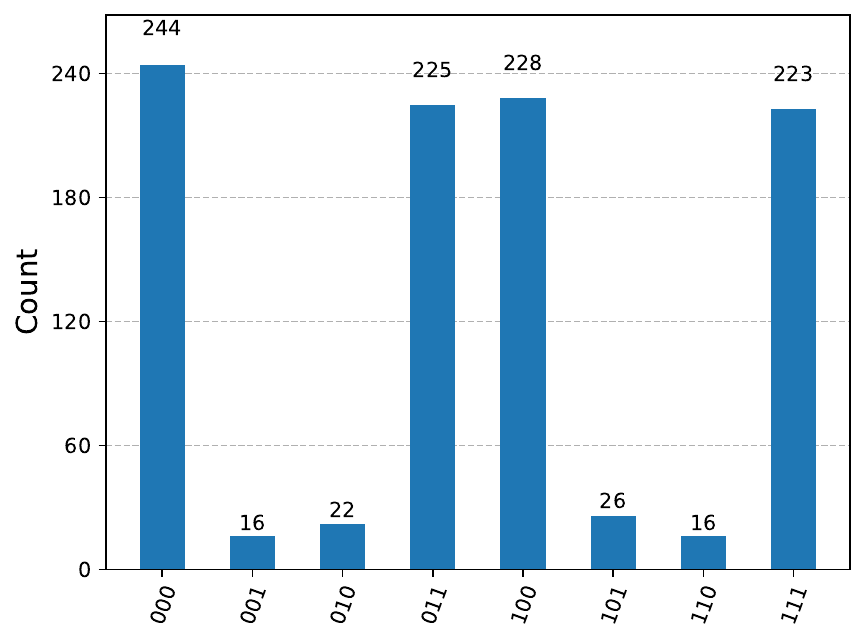}\label{fig:teleportation}}
    \hfill
    \subfloat[QFT Circuit]{\includegraphics[width=0.2\textwidth]{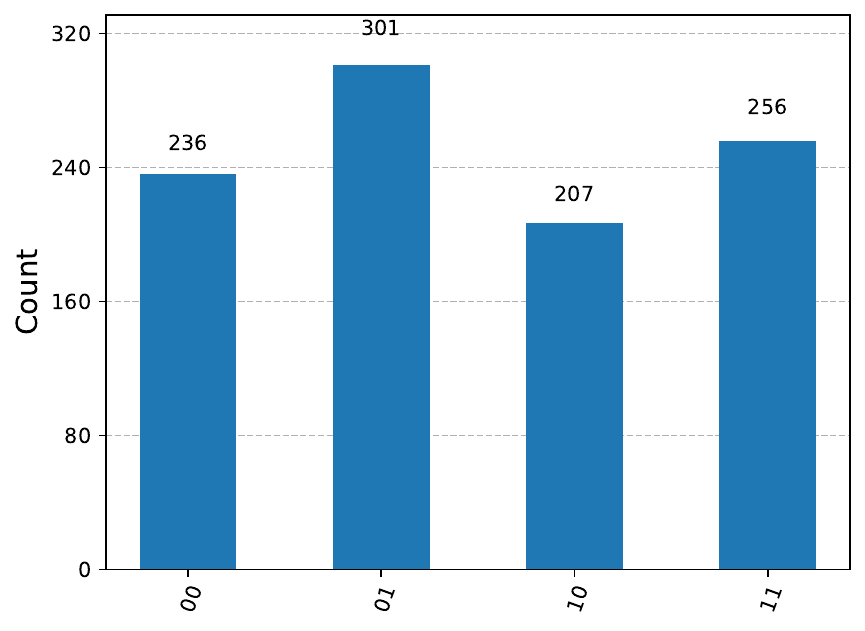}\label{fig:qft}}
    \hfill
    \subfloat[Bell State]{\includegraphics[width=0.2\textwidth]{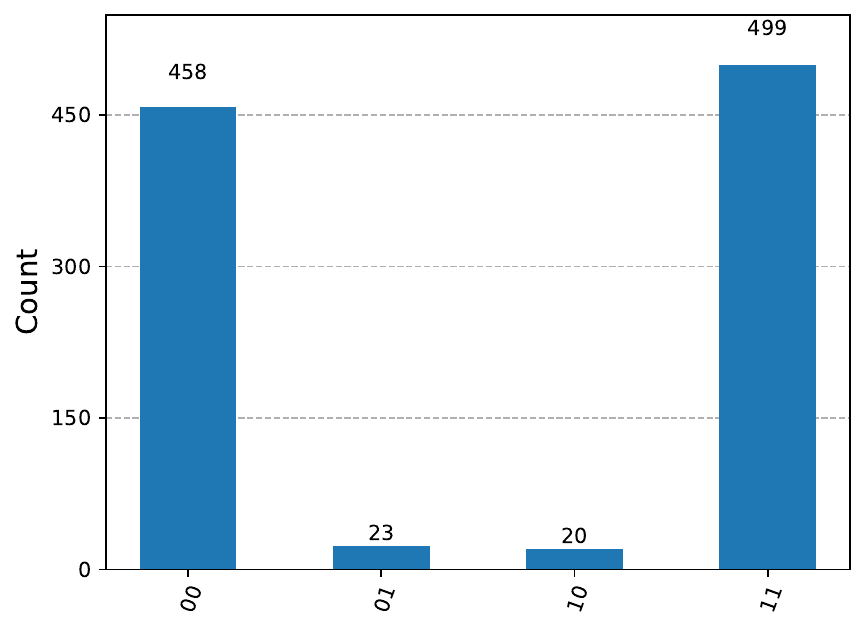}\label{fig:bell}}

    \caption{Quantum circuit reference outputs for the individual common user circuits.}
    \label{fig_single_benchmarks}
\end{figure*}

Table~\ref{tab_reset_scaling} shows how the cost per shot scales with different number of resets. Note that for the simple Bell state circuit, in terms of depth counted in number of gates, the circuit depth is $2$ so any number of resets more than $2$ means that the reset sequence is actually longer than the circuit itself. Consequently, the data in the table presents sort of worst case where cost is dominated by the number of resets so we can see their impact on the cost. For bigger circuits, the cost of resets will be less than cost of the circuit itself and the benefits would be better for the user. Even in this worst case, we can observe that around $2$ to $8$ resets gives significant savings to the user compared to executing each job separately.

\section{Common User Circuits Used For Extended Evaluation}

Before evaluating more complex examples, we now introduce $8$ common quantum computing circuits. These will be used to test the effectiveness of our exploit and how much free computation a malicious user can achieve with a mix of common circuits.

\subsection{Common User Circuits}

Common circuits that users may want to execute today include the following eight:

\begin{enumerate}
    \item \textbf{Bell State Circuit}: Generates two-qubit entanglement using a Hadamard gate followed by a CNOT. Outputs the state $\frac{1}{\sqrt{2}}(|00\rangle + |11\rangle)$.

    \item \textbf{Quantum Fourier Transform (QFT)}: The quantum analog of the discrete Fourier transform, composed of Hadamard and controlled phase gates. Used in Shor’s algorithm and phase estimation.

    \item \textbf{Quantum Teleportation Circuit}: Transfers an unknown quantum state using entanglement and classical communication. Involves Bell pair generation, entangling operations, and correction gates based on measurements.

    \item \textbf{Variational Quantum Circuit (Ansatz)}: Parametrized circuits used in VQE, QAOA, and other variational algorithms. Typically consists of layered rotations and entangling gates, optimized classically.

    \item \textbf{Grover’s Algorithm Circuit}: Implements quadratic speed-up for unstructured search problems. Uses an oracle and a diffusion operator applied iteratively.

    \item \textbf{Quantum Phase Estimation (QPE)}: Estimates the eigenvalue of a unitary operator. Employs controlled-unitary operations and QFT to extract phase information.

    \item \textbf{Bernstein–Vazirani Circuit}: Determines a hidden binary string with a single oracle query. Uses Hadamard gates before and after the oracle.

    \item \textbf{Deutsch–Jozsa Circuit}: Decides whether a given function is constant or balanced using a single query. Uses an oracle sandwiched between Hadamard layers.
\end{enumerate}

\begin{table*}[t]
\caption{Quantum circuit composite benchmarks using $4$ random common user circuits each.}
\centering
\begin{tabular}{cc}
\toprule
\textbf{Benchmark Mix} & \textbf{Component Circuits} \\
\midrule
Mix 4 A & Quantum Fourier Transform \quad Deutsch-Jozsa (Balanced) \quad Bell State \quad Quantum Teleportation \\
\hline
Mix 4 B & Quantum Teleportation \quad Deutsch-Jozsa (Balanced) \quad Phase Estimation \quad Variational Ansatz \\
\hline
Mix 4 C & Bernstein-Vazirani \quad Phase Estimation \quad Quantum Fourier Transform \quad Grover's Oracle \\
\hline
Mix 4 D & Bell State \quad Phase Estimation \quad Bernstein-Vazirani \quad Quantum Teleportation \\
\bottomrule
\end{tabular}
\label{tab_composite_benchmarks}
\end{table*}

\begin{figure*}[t]
    \centering

    \subfloat[Benchmark Mix 4 A]{\includegraphics[width=0.93\textwidth]{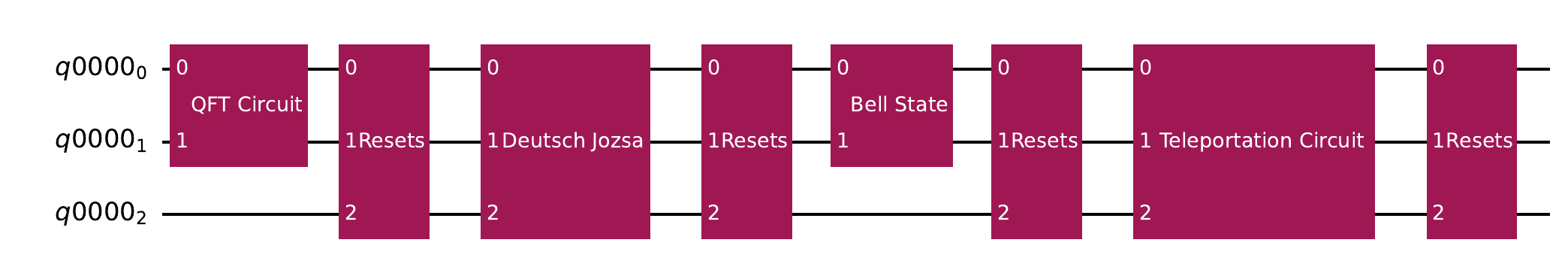}\label{fig_mix_1}}
    
    \vspace{0.3cm}
    
    \subfloat[Benchmark Mix 4 B]{\includegraphics[width=0.93\textwidth]{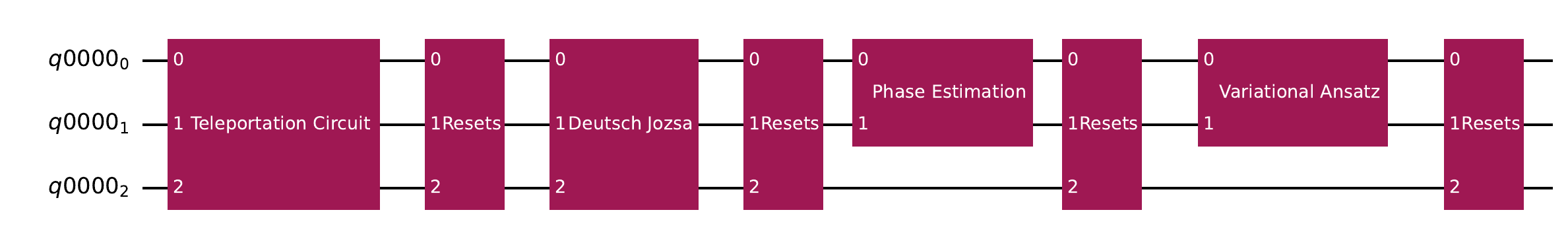}\label{fig_mix_2}}

    \vspace{0.3cm}
    
    \subfloat[Benchmark Mix 4 C]{\includegraphics[width=0.93\textwidth]{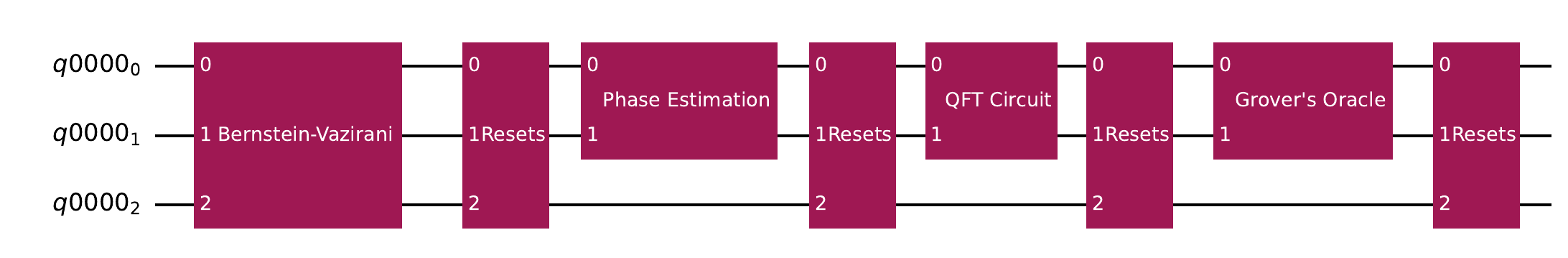}\label{fig_mix_3}}
        
    \vspace{0.3cm}
    
    \subfloat[Benchmark Mix 4 D]{\includegraphics[width=0.93\textwidth]{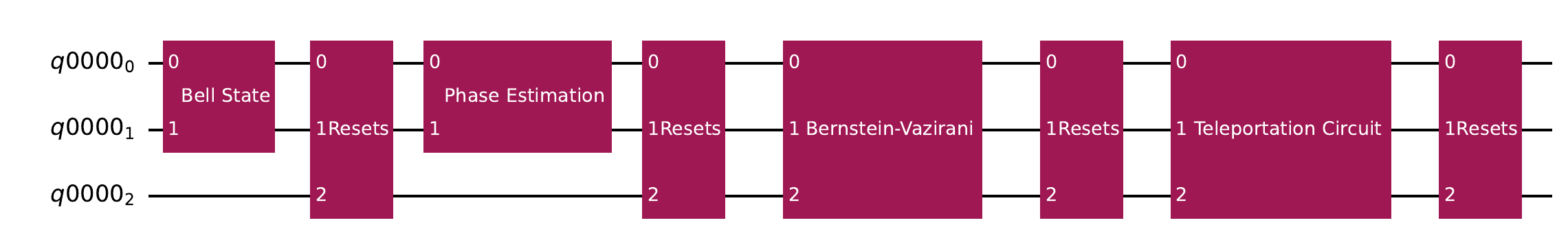}\label{fig_mix_4}}
    
    \caption{Composite benchmarks used.}
    \label{fig_composite_benchmarks}
\end{figure*}

\subsection{Execution of Common Circuits on a Quantum Computer Individually}
\label{sec_user_circuits_baseline}

We execute the common circuits individually on the target quantum computer. The purpose of this is to show the typical outputs of these circuits. The purpose is also to collect data about the outputs so that later we can use variational distance to compute how well individually executed circuits compare to when they are executed together as part of bigger circuit while being separated by resets. Due to limited quantum computing resources, we limit our evaluation of the user circuits to circuits in size from about $2$ to about $4$ qubits. Some circuits use ancilla qubits, so they may be slightly larger in size. 

Figure~\ref{fig_single_benchmarks} shows the reference outputs for each of the quantum circuits. The circuits were run individually, each for $1000$ shots. Each job was charged for $2$s of execution time, and cost $1.50$ credits on the target quantum computer.

\section{Executing Circuits for Free: Mixes of Common User Circuits}

We now demonstrate how users can run circuits for free in practical setting. In a realistic setting, a user may have a set of circuits they want to run. If they run circuits individually, they have to incur the per-task and per-job costs for each circuit. By combining multiple circuits into one job, and separating each copy of the circuit with $N$ resets, they can obtain computation for free as demonstrated in prior section. The scale of how much computation can be obtained for free in a practical scenario is reduced due to longer compilation times, and longer execution time. First, combining many (larger) circuits into one job creates an effectively larger circuit, which takes more compilation time. Second, because the effective circuit is larger, the per-shot execution time is longer incurring larger charges. Impact of these aspects is evaluated in this section.

Note that circuits of different widths (i.e. number of qubits) can be combined into one job. The size (width) of the combined circuit is simply the maximum of the number of number of qubits used by any one circuit. If circuits of different sizes (widths) are combined, the resets between each circuit will reset all the qubits between each copy of each circuit.

\subsection{Composite Benchmarks of 4 Circuits: Mix 4}

To evaluate realistic scenario of user trying to execute multiple, different circuits for free, we created a set of composite benchmarks. Each benchmark is created by randomly selecting $4$ of the circuits and concatenating them together into one bigger circuit that is submitted for execution. The circuits are selected randomly without regard to their width or depth. The final width of each of the composite benchmarks is the maximum width of any one of its component circuits. The depth depends on the total depth of each circuit plus the reset gates we insert between the circuits. We select $4$ resets as the number of resets that is inserted between each component circuit. The benchmark mixes used are listed in Table~\ref{tab_composite_benchmarks}.

Diagrams of the benchmarks used are shown in Figure~\ref{fig_composite_benchmarks}. It can be seen that not all component circuits use all qubits. Each reset box represents $4$ resets. The measurements and classical bits are not shown in the diagrams to save space.

\begin{figure*}[t]
    \centering

    \subfloat[QFT Circuit]{\includegraphics[width=0.2\textwidth]{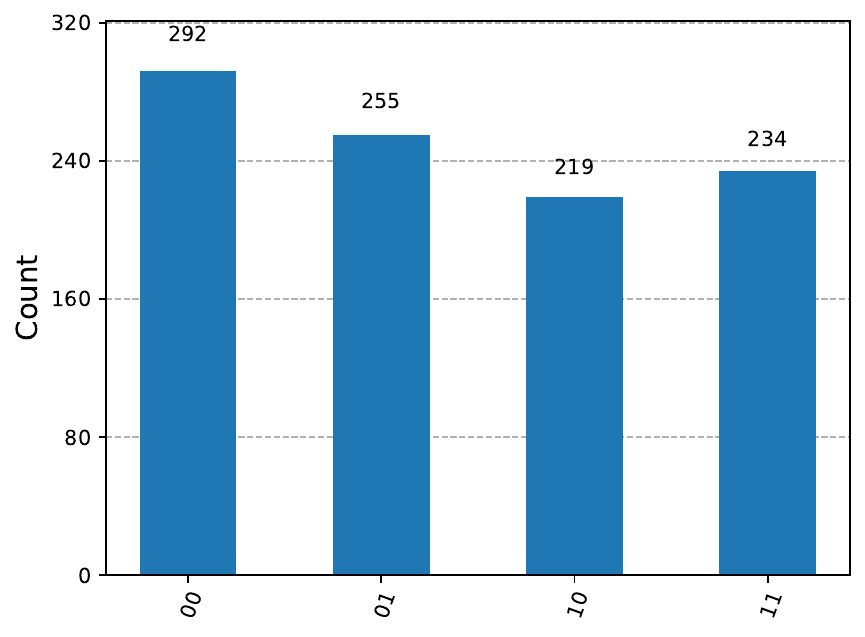}\label{fig:mix_1_qft_circuit_2q}}
    \hfill
    \subfloat[Deutsch-Jozsa]{\includegraphics[width=0.2\textwidth]{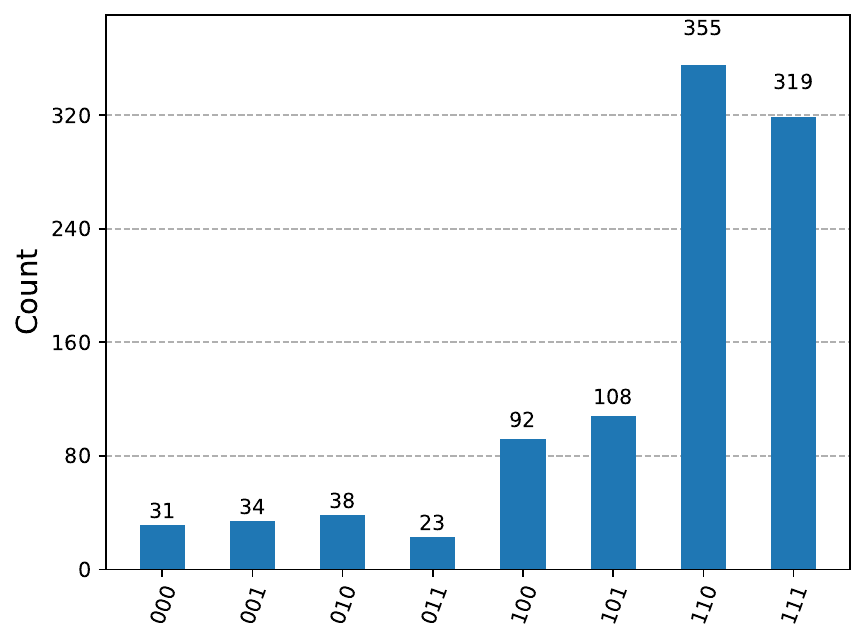}\label{fig:mix_1_deutsch_jozsa_balanced_2q}}
    \hfill
    \subfloat[Bell State]{\includegraphics[width=0.2\textwidth]{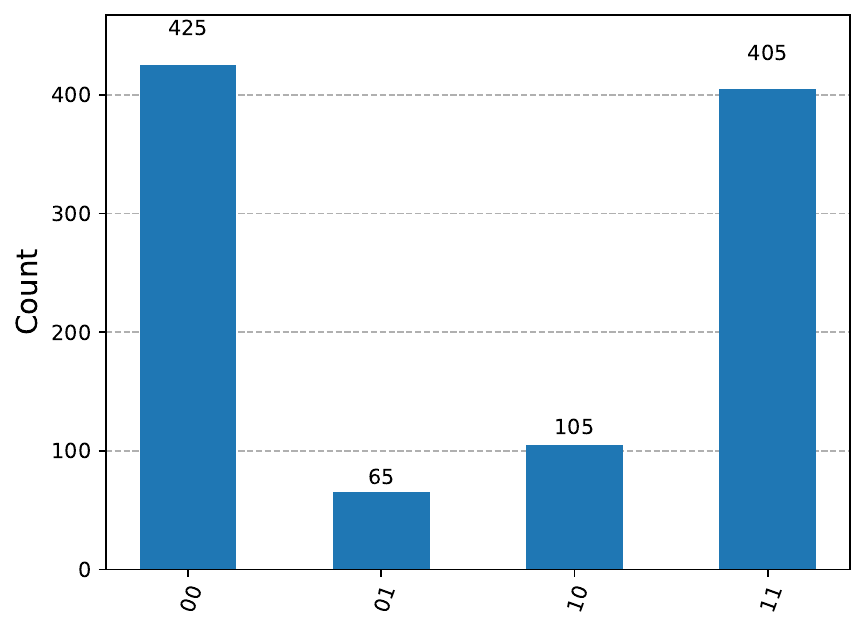}\label{fig:mix_1_bell_circuit_2q}}
    \hfill
    \subfloat[Teleportation]{\includegraphics[width=0.2\textwidth]{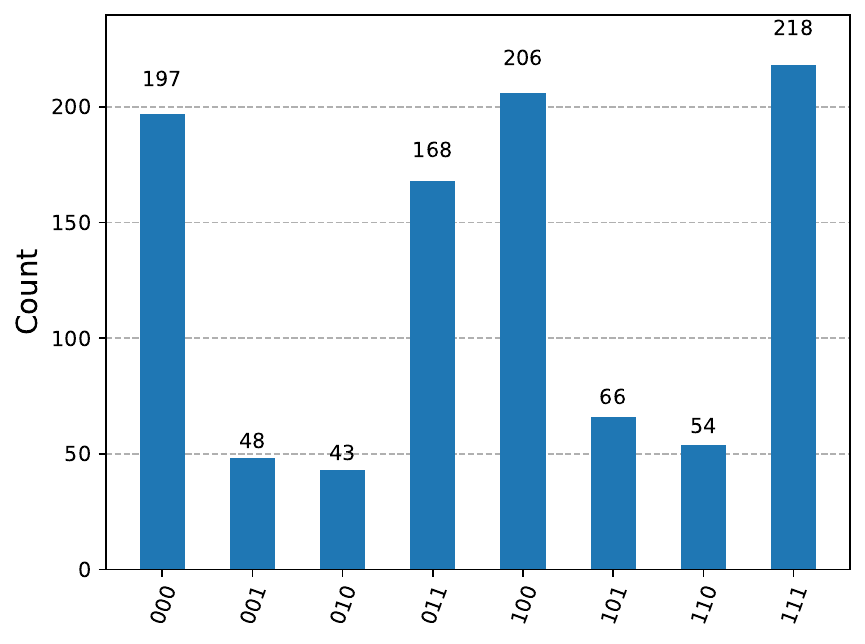}\label{fig:mix_1_teleportation_circuit_2q}}
    
    \caption{Composite benchmarks Mix 4 A results.}
    \label{fig_composite_mix_1_results}
\end{figure*}

\begin{figure*}[t]
    \centering

    \subfloat[Teleportation]{\includegraphics[width=0.2\textwidth]{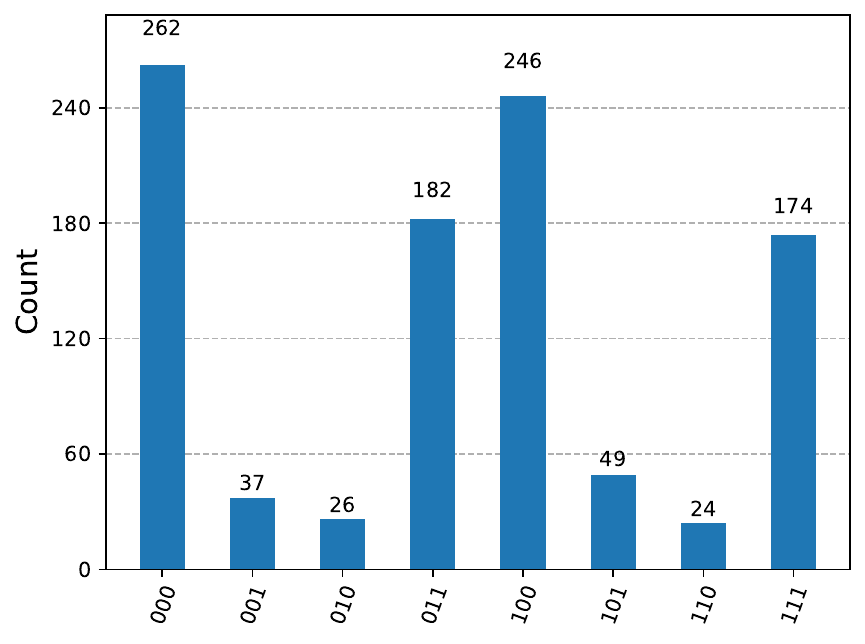}\label{fig:mix_2_teleportation_circuit_2q}}
    \hfill
    \subfloat[Deutsch-Jozsa]{\includegraphics[width=0.2\textwidth]{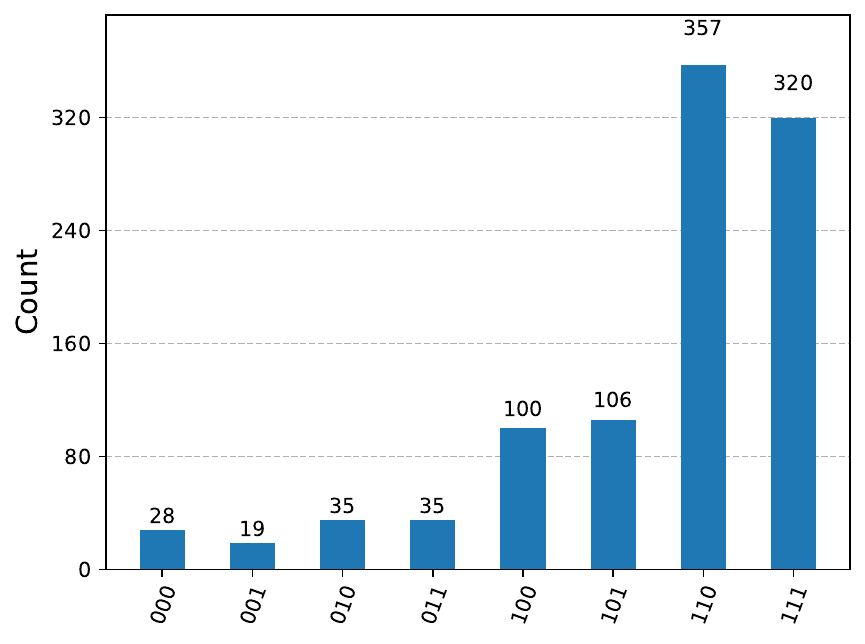}\label{fig:mix_2_deutsch_jozsa_balanced_2q}}
    \hfill
    \subfloat[Phase Estimation]{\includegraphics[width=0.2\textwidth]{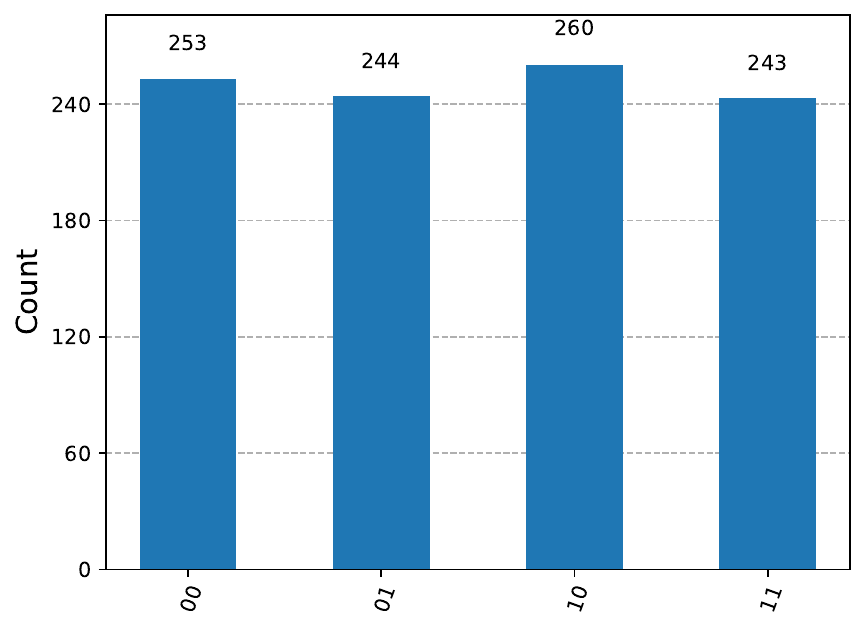}\label{fig:mix_2_phase_estimation_2q}}
    \hfill
    \subfloat[Variational Ansatz]{\includegraphics[width=0.2\textwidth]{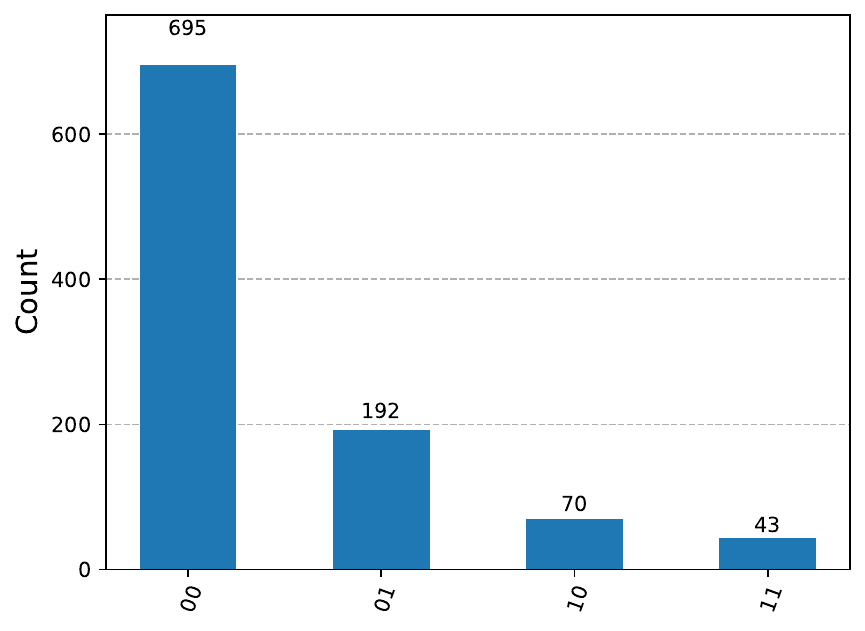}\label{fig:mix_2_variational_ansatz_2q}}
    
    \caption{Composite benchmarks Mix 4 B results.}
    \label{fig_composite_mix_2_results}
\end{figure*}

\begin{figure*}[t]
    \centering

    \subfloat[Bernstein-Vazirani]{\includegraphics[width=0.2\textwidth]{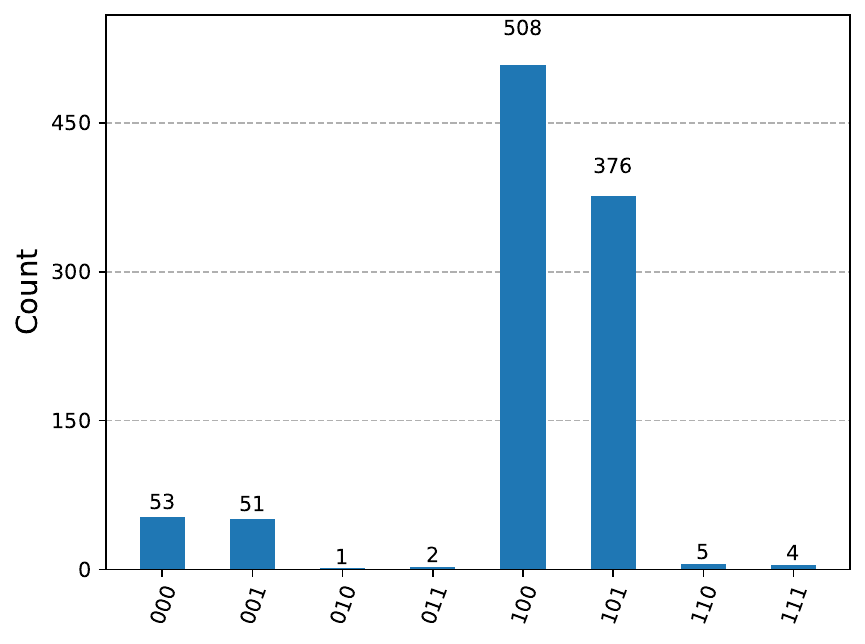}\label{fig:mix_3_bernstein_vazirani_2q}}
    \hfill
    \subfloat[Phase Estimation]{\includegraphics[width=0.2\textwidth]{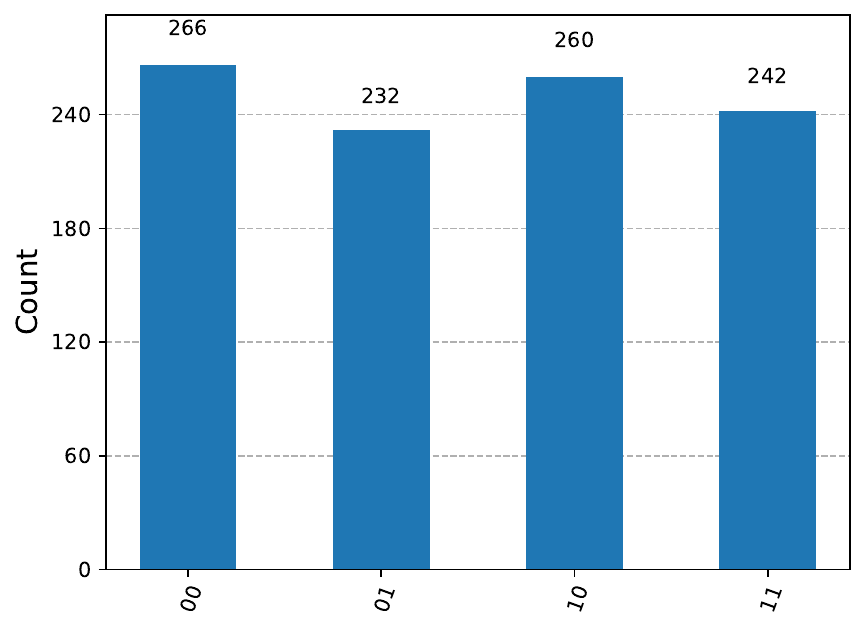}\label{fig:mix_3_phase_estimation_2q}}
    \hfill
    \subfloat[QFT Circuit]{\includegraphics[width=0.2\textwidth]{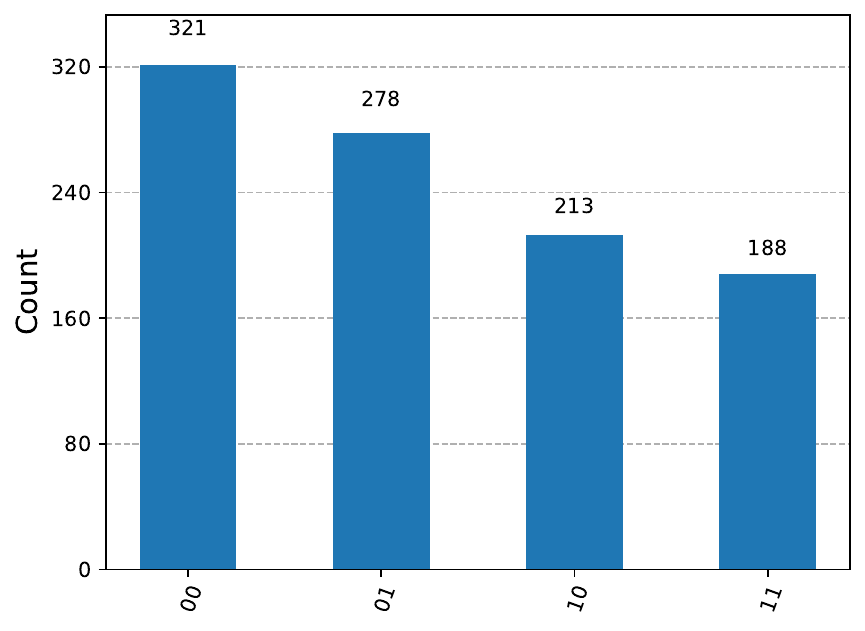}\label{fig:mix_3_qft_circuit_2q}}
    \hfill
    \subfloat[Grover's Oracle]{\includegraphics[width=0.2\textwidth]{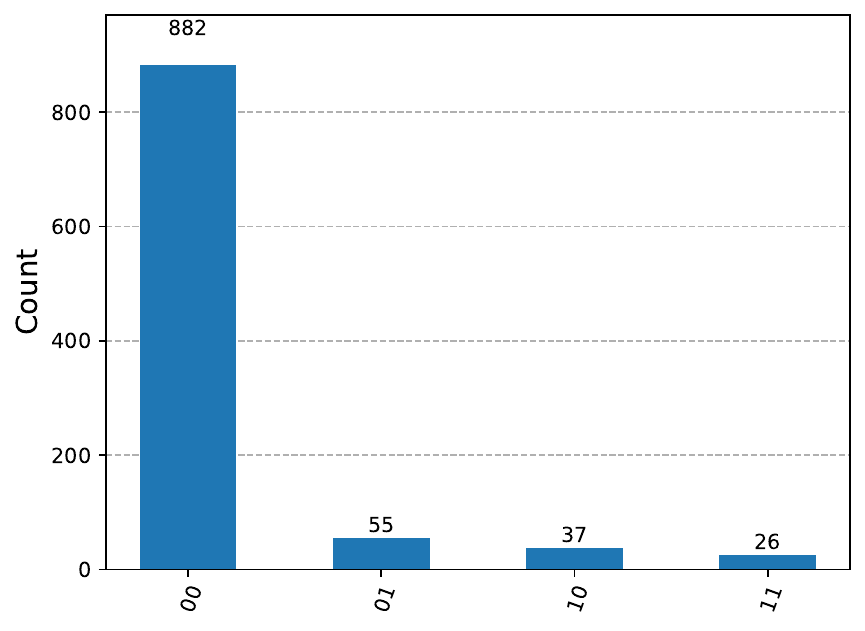}\label{fig:mix_3_grover_oracle_2q}}
    
    \caption{Composite benchmarks Mix 4 C results.}
    \label{fig_composite_mix_3_results}
\end{figure*}

\begin{figure*}[t]
    \centering

    \subfloat[Bell State]{\includegraphics[width=0.2\textwidth]{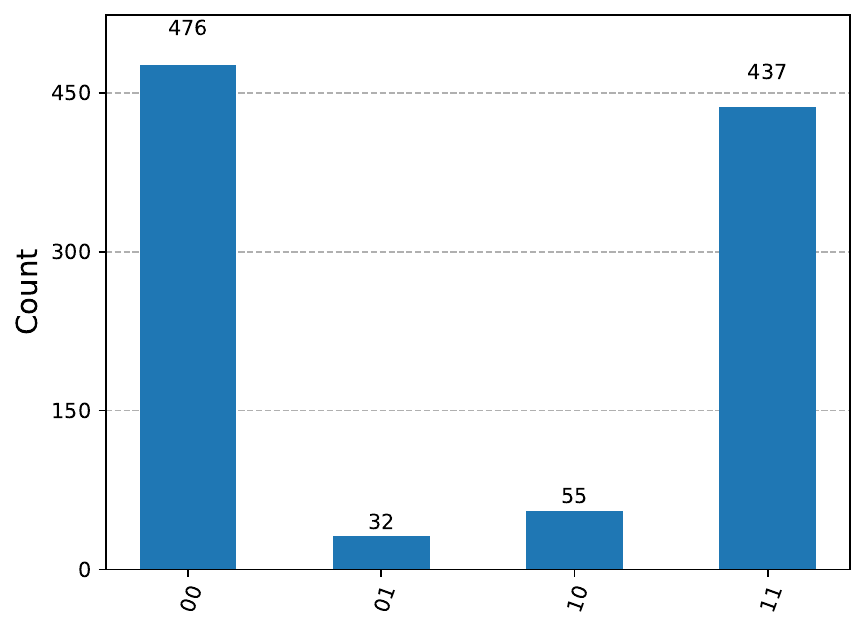}\label{fig:mix_4_bell_circuit_2q}}
    \hfill
    \subfloat[Phase Estimation]{\includegraphics[width=0.2\textwidth]{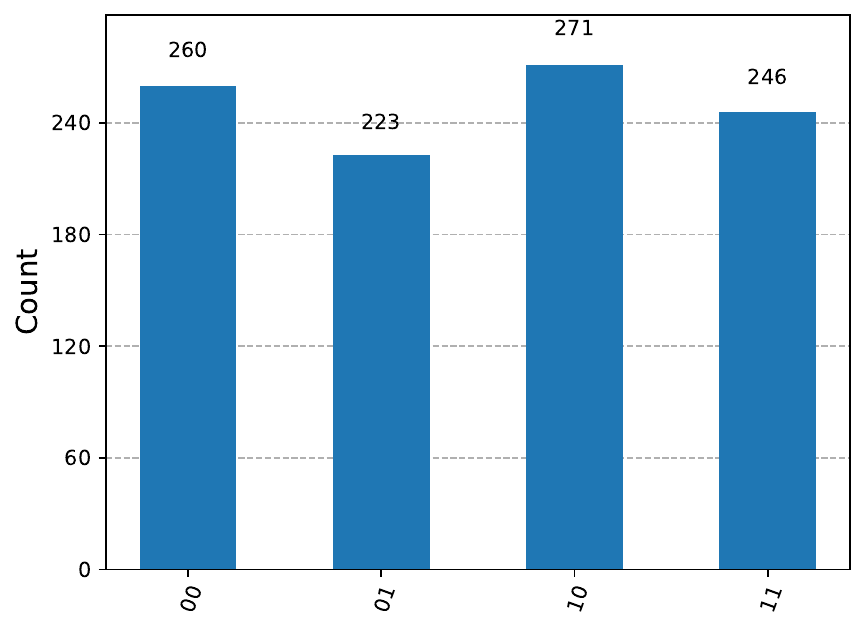}\label{fig:mix_4_phase_estimation_2q}}
    \hfill
    \subfloat[Bernstein-Vazirani]{\includegraphics[width=0.2\textwidth]{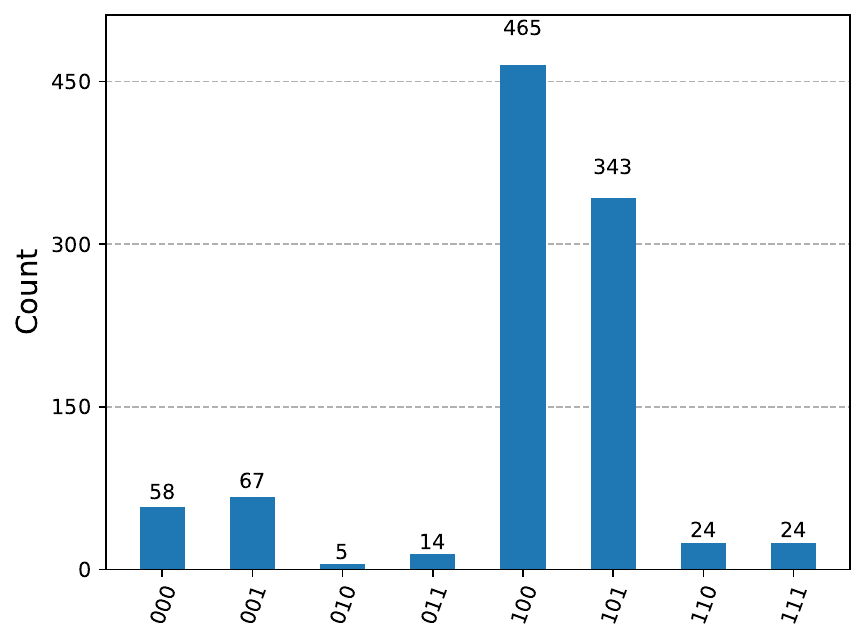}\label{fig:mix_4_bernstein_vazirani_2q}}
    \hfill
    \subfloat[Teleportation]{\includegraphics[width=0.2\textwidth]{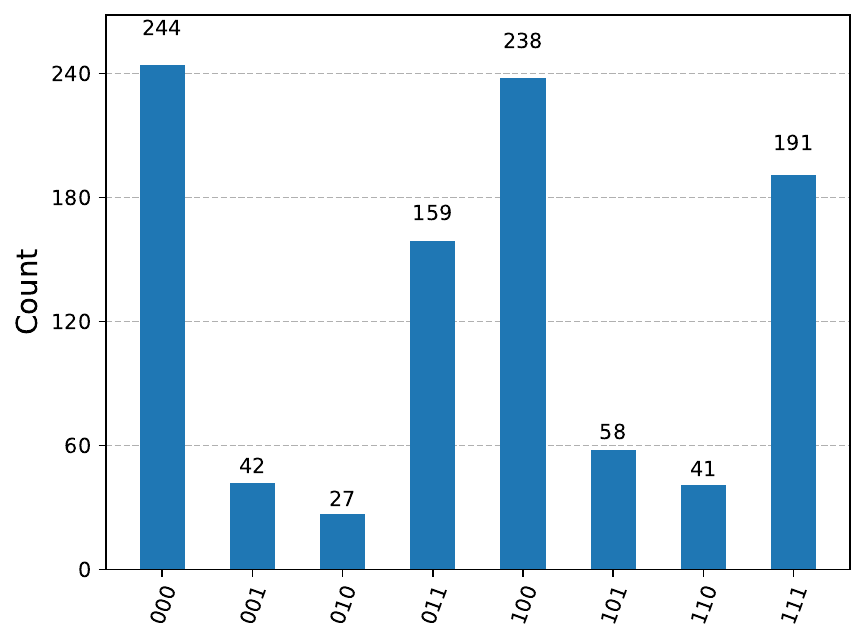}\label{fig:mix_4_teleportation_circuit_2q}}
    
    \caption{Composite benchmarks Mix 4 D results.}
    \label{fig_composite_mix_4_results}
\end{figure*}

\subsection{Free Computation Achieved for Mix 4 Benchmarks}

Table~\ref{tab_cost_savings_mix4} shows cost per shot, baseline cost per shot, and cost per shot savings for each benchmark in Mix 4 benchmarks. In subsequent section we try bigger mixes and show that the savings increase as the mix size increases.

\begin{table}[t]
\centering
\caption{Cost per shot, baseline cost per shot, and cost per shot savings for each benchmark in Mix 4 benchmarks.}
\label{tab_cost_savings_mix4}
\begin{tabular}{lccc}
\toprule
\textbf{Mix} & \textbf{Cost per Shot} & \textbf{Baseline Cost per Shot} & \textbf{Savings} \\
\textbf{Name} & \textbf{(credits)} & \textbf{(credits)} & \textbf{(\%)} \\
\midrule
Mix 4 A  & 0.000563 & 0.001500 & 166.67 \\
Mix 4 B  & 0.000563 & 0.001500 & 166.67 \\
Mix 4 C  & 0.000563 & 0.001500 & 166.67 \\
Mix 4 D  & 0.000563 & 0.001500 & 166.67 \\
\bottomrule
\end{tabular}
\end{table}

\subsection{Variational Distance of Benchmark Results}

\begin{table*}[t]
\centering
\caption{Amount of free computation achieved for different benchmark mixes. The user circuits used in each mix are shown in Table~\ref{tab_benchmarks_mix8_48} at the end of the paper. `Free Comp.' is the amount of free computation achieved, it is based on `Baseline Cost' minus `Cost'. Baseline cost is simply the number of effective shots times the baseline cost for each shot ($0.001500$).}
\label{tab_free_computation}
\begin{tabular}{l|c|c|c|c|c|c|c|c}
\hline
\textbf{Mix} & \textbf{Num. User Circuits} & \textbf{Shots} & \textbf{Eff. Shots} & \textbf{Time (s)} & \textbf{Cost} & \textbf{Baseline Cost} & \textbf{Free Comp.} & \textbf{Savings} \\
\hline
Mix 8  & 8   & 1000 & 8000  & 3  & 2.25  & 12   & 9.75  & 76.92\% \\
Mix 16  & 16  & 1000 & 16000 & 4  & 3.00  & 24   & 21.00 & 85.71\% \\
Mix 32  & 32  & 1000 & 32000 & 7  & 5.25  & 48   & 42.75 & 87.72\% \\
Mix 48  & 48  & 1000 & 48000 & 11 & 8.25  & 72   & 63.75 & 87.06\% \\
Mix 64  & 64  & 1000 & 64000 & 16 & 12.00 & 96   & 84.00 & 85.71\% \\
Mix 80  & 80  & 1000 & 80000 & 27 & 20.25 & 120  & 99.75 & 79.70\% \\
\hline
\end{tabular}
\end{table*}

\begin{table}[t]
\centering
\caption{Cost per shot, baseline cost per shot, and cost per shot savings for each benchmark in Mix 8 to Mix 80 benchmarks.}
\label{tab_cost_savings}
\begin{tabular}{lccc}
\toprule
\textbf{Mix} & \textbf{Cost per Shot} & \textbf{Baseline Cost per Shot} & \textbf{Savings} \\
\textbf{Name} & \textbf{(credits)} & \textbf{(credits)} & \textbf{(\%)} \\
\midrule
Mix 8   & 0.000281 & 0.001500 & 533.33 \\
Mix 16  & 0.000188 & 0.001500 & 800.00 \\
Mix 32  & 0.000164 & 0.001500 & 914.29 \\
Mix 48  & 0.000172 & 0.001500 & 872.73 \\
Mix 64  & 0.000188 & 0.001500 & 800.00 \\
Mix 80  & 0.000253 & 0.001500 & 592.59 \\
\bottomrule
\end{tabular}
\end{table}

\begin{table}[t]
\centering
\caption{Variational distances for each circuit within each benchmark mix. For example, for Mix 4 A, Circuit 1 is the QFT Circuit, Circuit 2 is the Deutsch-Jozsa circuit, etc. Please reference Table~\ref{tab_composite_benchmarks} for which circuits are within each benchmark.}
\label{tab_mix_outcomes_tvd}
\begin{tabular}{lcccc}
\toprule
\textbf{Mix} & \textbf{Circuit 1} & \textbf{Circuit 2} & \textbf{Circuit 3} & \textbf{Circuit 4} \\
\midrule
Mix 4 A & 0.068 & 0.179 & 0.127 & 0.131 \\
Mix 4 B & 0.092 & 0.182 & 0.075 & 0.043 \\
Mix 4 C & 0.069 & 0.087 & 0.091 & 0.090 \\
Mix 4 D & 0.062 & 0.085 & 0.613 & 0.617 \\
\bottomrule
\end{tabular}
\end{table}

We now compute the variational distances of the outputs of the circuits, compared to the baseline outputs from Section~\ref{sec_user_circuits_baseline}. Table~\ref{tab_mix_outcomes_tvd} shows the variational distances. We observe most circuits have variational distance $<0.1$ with only one outlier at $0.6$. However, note the outlier (Mix 4 D, Circuit 4) is same as Mix 4 A, Circuit 4, where the same circuit at the same position in the mix (last circuit) has variational distance of $0.131$. These large changes are unfortunate side-effect of today's NISQ quantum computers, they are not due to our strategy of placing many circuits together separated by resets. We conclude that many circuits executed together within larger benchmark give results no worst than individually executed circuits, considering inherent noise in NISQ quantum computers.

\subsection{Composite Benchmarks of 8 to 80 Circuits: Mix 8 to Mix 80}

In this section we further evaluate how much free computation an attacker could achieve when running different mixes of user circuits. Especially, we increase the number of user circuits in each benchmark mix and evaluate mixes that have from $8$ to $80$ user circuits in them. For each mix, the user circuits were randomly selected from among the common user circuits to build the benchmark mix. Python {\tt random.choices()} to generate the list of randomly selected user circuits to include in each mix. The resulting user circuits used in each mix are listed in Table~\ref{tab_benchmarks_mix8_48} at the end of the paper.

For each benchmark mix, as before, we used $1000$ shots and $4$ resets between each user circuit within the benchmark mix circuit. Because multiple circuits are executed within one shot, the number of effective shots is $1000$ times the number of user circuits in the mix. Due to the size of the circuits, the circuit diagrams and the output probabilities are not shown.

\subsection{Free Computation Achieved for Mix 8 to Mix 80 Benchmarks}

Each benchmark mix was executed on the target quantum computer and the time (s) and cost (credits) was recorded as shown in Table~\ref{tab_free_computation}. From this data, we computed how much it would cost to run the user circuits in each benchmark if they were run individually (shown in `No Attack' cost column) and how much it costs when circuits are combined with resets between them (shown in 'With Attack' cost column, which is same as the `Cost' column). The amount of free computation is shown in the `Free Comp.' cost column. It can be seen that users can achieve $10$s of credits in free computation and savings are in the range of $80$\%.

We also investigate cost per shot in Table~\ref{tab_cost_savings}. We compute the cost per shot as the cost divided by the number of effective shots. The baseline cost per shot is based on just running one user circuit for $1000$ shots which typically costs $1.5$ credits thus as result baseline cost per shot is $1.5 / 1000 = 0.001500$ credits. It can be observed from the table that around $32$ user circuits is where the cost per shot is the lowest and a malicious user is able to achieve the highest ratio of free computation. This correlates also with Table~\ref{tab_bell_cost_scaling} where around $32$ copies of Bell state circuit also gave the highest percentage for saving on per shot costs.

\section{Mitigation and Defense}
\label{sec_defense}

Our results demonstrate that current billing mechanisms in cloud-based quantum computing platforms, when combined with the availability of mid-circuit measurement and reset operations, enable a malicious user to obtain effectively free computation. In particular, per-task + per-shot and time-based pricing models allow an attacker to concatenate many independent circuits into a single task and thereby amortize or eliminate the cost that is intended to scale with computational workload. Therefore, the most effective mitigation is conceptually simple: cloud providers should avoid billing schemes that depend on the number of shots or the duration of execution. Instead, providers should adopt gate-based billing, where users are charged according to the number of gates in the submitted~circuit.

Gate-based billing aligns closely with the physical resources consumed on the device. After transpilation, the cloud provider knows exactly which gates appear in the circuit, how many instances of each gate type are present, and the expected duration of each operation. This information is already required for scheduling, compilation, and hardware calibration, and thus introduces no new overhead. Under this model, resets naturally incur a cost proportional to the operations required to implement them, and users cannot bypass billing by embedding multiple user-level circuits into one larger circuit. Importantly, this approach preserves compatibility with all standard workflows and does not require detecting user intent or analyzing circuit semantics.

Alternative mitigations are substantially weaker. One approach is to disable mid-circuit reset operations entirely. However, resets are a critical feature in modern quantum devices: they enable circuit depth reduction, dynamic circuit architectures, syndrome extraction in error-correcting codes, and algorithmic primitives such as iterative phase estimation. Removing this functionality would degrade performance for benign users and restrict the expressiveness of hardware that already supports resets. Consequently, blanket disabling of resets is not a viable or acceptable solution.

A different class of defenses attempts to detect malicious circuit concatenation behavior. For example, a provider might flag circuits in which all qubits are reset at some point, or circuits whose structure appears to repeat multiple times within a single shot. However, such detection is fundamentally heuristic and can be evaded. An adversary could stagger resets across subsets of qubits, reset in patterns that mimic normal algorithmic structures, or exploit properties of common techniques such as uncomputation. Uncomputation, which reverses intermediate operations to return ancilla qubits to the $\ket{0}$ state, is widely used in algorithms such as HHL, amplitude estimation, and arithmetic routines. Circuits that legitimately use uncomputation naturally contain qubits that appear to “reset themselves” without explicit reset gates, making it difficult to distinguish malicious concatenation from valid algorithmic behavior. As a result, detection-based defenses risk both false positives on legitimate workloads and false negatives on adversarial ones.

Given these limitations, gate-based billing offers the most robust and principled mitigation. It directly ties the user’s cost to the actual quantum resources consumed, removes the economic incentive for abuse, and preserves the full expressiveness of mid-circuit measurement and reset operations. By adopting this model, cloud providers can eliminate the attack vector identified in this work while maintaining usability and compatibility with existing quantum programming frameworks.

\section{Related Work}
\label{sec:conclusion}

The security of quantum computers, distinct from classical research in post-quantum cryptography, is an emerging area of study, with relatively limited existing literature for comparison.

\subsection{Attacks on Reset Gates}

Considering reset gates and operations, prior work by Mi et al.~\cite{mi2022securing} proposed an attack on the reset operations. The authors showed how it is possible to recover (some) classical information across reset gates, leading to a type of side-channel attack. Later work extended this prior research and showed more advanced attacks where use of concealing circuit is used to help hide the attacker while still allowing for information leak to be extracted by the attacker across reset gates. Besides, insecure reset is also a problem when considering lower-level support, such as the higher-energy attacks proposed in~\cite{10.1145/3576915.3623104}, which demonstrated that attackers can abuse higher-energy states to bypass normal quantum gates and operations. 

In our work, we do not deal with higher-energy attacks nor side-channels. We simply use resets a means to reset qubits so that multiple circuits can execute within one shot, while the user is only billed for a shot and not all the circuits.

\subsection{Defenses for Attacks on Reset Gates}

Considering the threat of reset gate attacks, prior work~\cite{mi2022securing} proposed defenses based on a new type of reset gate. More recent work~\cite{xu2025securing} has presented a new one-time pad approach for randomizing the state of the qubits after the reset operations.

In our work, we are not concerned with attackers bypassing the reset operations to leak information because all the attacker's circuits are run together. The attacker may actually benefit from the existing defenses such as~\cite{xu2025securing} as they will improve the fidelity of the qubits reducing the noise and errors when many circuits are run in one shot while being separated by the (now improved) reset gates.

\subsection{Crosstalk Attacks}

NISQ quantum computers are prone to errors and noise from many sources. Besides decreasing the fidelity of the quantum computing programs' results, these errors and noise may also be taken advantage of by attackers to perform malicious attacks or retrieve secret information. The crosstalk among quantum gates and qubits has been researched and shown that quantum jobs in a multi-tenant quantum computing architecture are vulnerable to interference from circuits that execute in parallel~\cite{10.1145/3370748.3406570, 9193969, 9840181, 10133711, deshpande2023design}. Such attacks are realized by paralleling quantum circuits, with one of the circuits generating a large crosstalk effect and interfering with the other circuit that is running at the same time. 
For example, malicious users can exploit crosstalk in various ways. For instance, {\tt CNOT} gates can induce significant crosstalk, allowing attackers to design circuits that disrupt nearby qubits instead of performing valid computations~\cite{harper2024crosstalk}. 
Furthermore, crosstalk can be used to target specific quantum algorithms, such as ensuring that Grover’s algorithm consistently returns incorrect results, thereby sabotaging more complex quantum jobs~\cite{deshpande2023design}. 

In our work we do not attack other circuits, but instead focus on exploiting and abusing the reset gates to gain free computation on quantum computers hosted in the cloud and billed on per-task plus per-shot basis, or on time-based~basis.

\subsection{Defenses for Crosstalk Attacks}

On the defense side, previous work has suggested an ``antivirus'' program that can be used to detect malicious quantum circuits~\cite{9840181,deshpande2023design}. Such antivirus could be used to help find instances of reset gates. However, as with any antivirus, attackers could find different or unusual ways of arranging the resets that may bypass the patterns that the antivirus is looking~for.

\section{Conclusion}

This work presented the first thorough exploration of how reset operations in cloud-based quantum computers could be exploited to run quantum circuits for free. Our evaluation on real cloud-based quantum computers demonstrated that the per-shot cost of executing certain circuits could be reduced by up to nearly 900\%, highlighting a previously unrecognized vulnerability with significant potential financial implications for quantum computing service providers. To address this novel security and billing concern, we proposed an updated approach for user billing based on per-gate execution rather than per-task, per-shot, or time-based methods. The per-gate billing mechanism ensures that users are charged precisely for the quantum operations they execute within each shot, preserving the flexibility and usability of mid-circuit measurements and active reset operations while mitigating the risk of abuse. Beyond the immediate mitigation, our findings have broader implications for the security and economics of cloud-based quantum computing. They underscore the importance of aligning billing strategies with the underlying physical operations to prevent unintended exploitation. Additionally, this work highlights that as quantum computers continue to scale, subtle features such as mid-circuit resets, intended to improve efficiency, can introduce novel attack vectors if not carefully managed.

\section{Responsible Disclosure}
\label{sec_disclosure}

We have disclosed the vulnerability to the affected quantum computer provider in July 2025. They have acknowledged the receipt of our early paper draft, but did not comment on the vulnerability itself. We believe that after over 3 months, sufficient responsible disclosure time has passed to discuss our finds publicly. We believe that by exposing this vulnerability, proposing a solution, and notifying affected quantum computer provider we can improve the security of the systems and prevent this vulnerability from becoming wide-spread.

\bibliographystyle{IEEEtran}
\bibliography{bibliography.bib}

\balance

\appendices 

\section{Full List of User Circuits}
\label{sec_appendix}

Table~\ref{tab_benchmarks_mix8_48} on next page shows full list of user circuits used in each benchmark mix for mixes Mix 8 to Mix 48. The names of the user circuits are listed in the order in which they appear in the corresponding benchmark mix.

\begin{table}[ht]
\centering
\caption{List of benchmark mixes.}
\label{tab_benchmarks_mix8_48}
\begin{tabular}{|p{0.6cm}|p{7cm}|}
\hline
\textbf{Mix} & \textbf{Order of Circuits in the Mix} \\
\hline
Mix 8 &
GHZ, Bernstein–Vazirani, Bell, Teleportation, Grover, Deutsch–Jozsa (Balanced), Phase Estimation, Variational Ansatz \\
\hline
Mix 16 &
Teleportation, Variational Ansatz, Bernstein–Vazirani, Bernstein–Vazirani, Variational Ansatz, Teleportation, Teleportation, Bell, Bernstein–Vazirani, Bernstein–Vazirani, Bell, Bernstein–Vazirani, Bernstein–Vazirani, Grover, Grover, Bernstein–Vazirani \\
\hline
Mix 32 &
Bernstein–Vazirani, Deutsch–Jozsa (Balanced), Deutsch–Jozsa (Balanced), Teleportation, Bernstein–Vazirani, Phase Estimation, Variational Ansatz, Deutsch–Jozsa (Balanced), Teleportation, Bell, Grover, Bell, Phase Estimation, Phase Estimation, Phase Estimation, Deutsch–Jozsa (Balanced), Bernstein–Vazirani, Variational Ansatz, Teleportation, Teleportation, Grover, Bell, Bell, Deutsch–Jozsa (Balanced), Variational Ansatz, Teleportation, Variational Ansatz, Bernstein–Vazirani, Grover, Teleportation, Bernstein–Vazirani, Grover \\
\hline
Mix 48 &
Variational Ansatz, Bell, Bernstein–Vazirani, Bernstein–Vazirani, Phase Estimation, Grover, Bernstein–Vazirani, Bell, Bernstein–Vazirani, Bell, Bernstein–Vazirani, Phase Estimation, Teleportation, Bernstein–Vazirani, Teleportation, Teleportation, Deutsch–Jozsa (Balanced), Teleportation, Variational Ansatz, Teleportation, Phase Estimation, Teleportation, Grover, Grover, Phase Estimation, Teleportation, Variational Ansatz, Grover, Teleportation, Grover, Bernstein–Vazirani, Teleportation, Variational Ansatz, Bell, Teleportation, Deutsch–Jozsa (Balanced), Bernstein–Vazirani, Bell, Bell, Grover, Phase Estimation, Teleportation, Bell, Bernstein–Vazirani, Grover, Grover, Grover, Deutsch–Jozsa (Balanced) \\
\hline

Mix 64 &
Bell, Variational Ansatz, Phase Estimation, Teleportation, Bernstein–Vazirani, Variational Ansatz, Phase Estimation, Variational Ansatz, Phase Estimation, Phase Estimation, Phase Estimation, Grover, Deutsch–Jozsa (Balanced), Grover, Teleportation, Phase Estimation, Deutsch–Jozsa (Balanced), Teleportation, Bernstein–Vazirani, Phase Estimation, Grover, Deutsch–Jozsa (Balanced), Teleportation, Variational Ansatz, Deutsch–Jozsa (Balanced), Bell, Bernstein–Vazirani, Teleportation, Teleportation, Grover, Bernstein–Vazirani, Bernstein–Vazirani, Teleportation, Teleportation, Phase Estimation, Phase Estimation, Deutsch–Jozsa (Balanced), Phase Estimation, Deutsch–Jozsa (Balanced), Deutsch–Jozsa (Balanced), Variational Ansatz, Bernstein–Vazirani, Bell, Teleportation, Teleportation, Phase Estimation, Bell, Grover, Deutsch–Jozsa (Balanced), Phase Estimation, Deutsch–Jozsa (Balanced), Deutsch–Jozsa (Balanced), Teleportation, Phase Estimation, Bernstein–Vazirani, Bell, Deutsch–Jozsa (Balanced), Phase Estimation, Variational Ansatz, Teleportation, Bell, Deutsch–Jozsa (Balanced), Grover, Phase Estimation \\
\hline
Mix 80 &
Teleportation, Phase Estimation, Bell, Deutsch–Jozsa (Balanced), Grover, Bell, Variational Ansatz, Phase Estimation, Bell, Phase Estimation, Phase Estimation, Phase Estimation, Deutsch–Jozsa (Balanced), Teleportation, Grover, Grover, Phase Estimation, Phase Estimation, Variational Ansatz, Teleportation, Variational Ansatz, Teleportation, Deutsch–Jozsa (Balanced), Bernstein–Vazirani, Phase Estimation, Teleportation, Bell, Bernstein–Vazirani, Teleportation, Phase Estimation, Phase Estimation, Phase Estimation, Deutsch–Jozsa (Balanced), Deutsch–Jozsa (Balanced), Teleportation, Deutsch–Jozsa (Balanced), Deutsch–Jozsa (Balanced), Deutsch–Jozsa (Balanced), Bernstein–Vazirani, Deutsch–Jozsa (Balanced), Bernstein–Vazirani, Grover, Grover, Phase Estimation, Teleportation, Teleportation, Phase Estimation, Grover, Grover, Phase Estimation, Phase Estimation, Bernstein–Vazirani, Teleportation, Grover, Teleportation, Bernstein–Vazirani, Variational Ansatz, Variational Ansatz, Bell, Variational Ansatz, Bell, Teleportation, Grover, Grover, Grover, Teleportation, Bell, Grover, Phase Estimation, Teleportation, Bernstein–Vazirani, Bell, Bernstein–Vazirani, Bernstein–Vazirani, Variational Ansatz, Teleportation, Bernstein–Vazirani, Deutsch–Jozsa (Balanced), Grover, Phase Estimation \\
\hline
\end{tabular}
\end{table}

\end{document}